\documentclass[pdflatex,sn-mathphys]{sn-jnl}

\jyear{2026}

\usepackage[T1]{fontenc}
\usepackage[utf8]{inputenc}
\usepackage{enumitem}
\usepackage{microtype}
\catcode`\|=12
\usepackage{tikz}
\usetikzlibrary{shapes.geometric, arrows.meta, positioning, fit, backgrounds, calc}
\usepackage{xcolor}

\newcommand{\acceptancenote}{%
  Accepted for publication in \emph{Machine Intelligence Research} (MIR), Springer Nature, 2026.}
\hypersetup{
  pdftitle={A Visionary Look at Vibe Researching},
  pdfauthor={Yebo Feng and Yang Liu},
}

\title[A Visionary Look at Vibe Researching]{A Visionary Look at Vibe Researching}

\author*[1]{\fnm{Yebo} \sur{Feng}}\email{yebo.feng@ntu.edu.sg}
\author[1]{\fnm{Yang} \sur{Liu}}\email{yangliu@ntu.edu.sg}

\affil[1]{\orgname{Nanyang Technological University}, \orgaddress{\city{Singapore}, \country{Singapore}}}

\abstract{As large language models (LLMs) reshape knowledge work, the notion of ``vibe coding'' has captured the software engineering zeitgeist: developers describe intent in plain English and let AI write the code. A parallel shift appears to be taking shape in scientific research. Under the banner of \textbf{vibe researching}, researchers delegate much of the mechanical burden of their work (literature surveys, coding, data wrangling, drafting) to LLM-based agents while retaining intellectual control. This forward-looking perspective article characterizes vibe researching as an emerging, cross-domain research paradigm. We trace its roots in vibe coding, distinguish it from ``AI for Science'' and fully autonomous ``auto research'' systems, and outline a conceptual framework for how it operates: multi-agent architectures, memory mechanisms, tool use, planning, retrieval-augmented generation, self-reflection, a reference architecture that names the components a vibe-researching system needs, and the human's role as creative director. We discuss eight technical limitations (hallucination, context-window constraints, infrastructure gaps, limited multimodal capability, verification asymmetry, brittleness on novel tasks, data privacy, and prompt-injection and adversarial retrieved content) and consider societal impacts on both sides: positive (broader access, faster iteration, expanded coverage, easier cross-disciplinary work, reduced cognitive load, surfaced hidden connections) and negative (convergent thinking, unsettled credit norms, literature flooding, polished mediocrity, eroded trust, devalued expertise, weakened training), and propose concrete reconciliation strategies for the most consequential tensions. For each problem, we sketch concrete future directions with possible technical approaches, including controlled empirical evaluation of vibe researching itself and dedicated agent-security defenses against adversarial corpora. Our goal is to offer the community a clearer map of the territory so that discussion and adoption can proceed from shared ground.
}


\makeatletter
\def\ps@titlepage{%
  \def\@oddhead{\vbox to 0pt{\hbox to \hsize{\hfil\footnotesize\acceptancenote\hfil}}}%
  \let\@evenhead\@oddhead%
  \def\@oddfoot{\vbox to 18pt{\vfill\reset@font\rmfamily\hfil\thepage\hfil}}%
  \def\@evenfoot{}}
\def\ps@headings{%
  \def\@evenhead{\hbox to \hsize{\headerfont\thepage\qquad\rightmark\hfill}}%
  \def\@oddhead{\hbox to \hsize{\headerfont\hfill\leftmark\qquad\thepage}}%
  \let\@mkboth\markboth}
\makeatother

\begin{document}

\maketitle

\section{Introduction}

Scientific research has always run on a peculiar combination of creativity and drudgery. The creative part, asking a question nobody has thought to ask, seeing a pattern in data that others overlooked, connecting ideas from fields that do not normally talk to each other, is what draws people into research and what produces the breakthroughs that move fields forward. The drudgery, reading hundreds of papers to map a literature, reimplementing someone else's baseline because they did not release their code, cleaning datasets, formatting manuscripts, chasing down a bug in an experimental pipeline at 2\,a.m., is what fills most of a researcher's actual working hours. For as long as science has existed, these two sides have been bundled together: you could not do the creative work without also doing the mechanical work, and the mechanical work was widely understood to be part of how researchers develop the intuition and expertise that makes the creative work possible.

Large language models are unbundling them. Models like GPT-4 \citep{openai2023gpt4} and Claude \citep{anthropic2024claude} can now write functional code, summarize papers, draft manuscript sections, design experimental setups, and carry on multi-turn technical conversations that stay coherent over dozens of exchanges. They are not perfect at any of these tasks, but they are good enough to be useful, and they are improving fast. The research community has noticed. The question is no longer whether AI will change how research gets done, but how, how much, and under whose control.

Two visions have crystallized. One pushes toward full automation. Systems like The AI Scientist \citep{lu2024aiscientist} aim to close the entire loop: generating hypotheses, writing and running code, producing figures, drafting complete papers, and even simulating peer review, all without human involvement. The ambition is to turn scientific discovery into a scalable computational process. The results so far are genuinely impressive in scope, though independent evaluations have noted significant failure rates and quality concerns \citep{schmidgall2025agentlab}. The other vision keeps the human firmly in charge, treating AI as a powerful but subordinate tool, not unlike how researchers already use statistical software or simulation packages, just far more general-purpose.

Between these poles, we argue that a third pattern is becoming visible, one that borrows its name and some of its ethos from the software world. In February 2025, Andrej Karpathy coined the term \textit{vibe coding} \citep{karpathy2025vibecoding} to describe a programming style where the developer expresses intent in natural language and lets the model generate the code. The developer talks; the model codes; the developer iterates by feel rather than by reading diffs. The term circulated widely (Section~\ref{sec:vibe_coding}) because it named a workflow many developers recognized. Karpathy later pushed the idea further with AutoResearch \citep{karpathy2025autoresearch}, a project that lets AI agents autonomously modify training code, run experiments, and iterate on results overnight. The underlying logic of vibe coding, express what you want in plain language, let the machine handle execution, and steer by evaluating outputs rather than producing them, appears to apply beyond software as well.

\textit{Vibe researching} extends this logic to the research setting. The researcher supplies the question, the domain intuition, and the critical eye. LLM-based agents handle much of the labor-intensive execution: searching and synthesizing literature, implementing methods, running experiments, processing data, generating figures, and drafting text. The researcher operates at the level of strategic direction and quality judgment rather than mechanical execution, intervening to steer, evaluate, and course-correct while delegating substantial portions of the work. It is not a formal methodology introduced by a single prior paper. Rather, we use the term to describe an emerging pattern that has become more recognizable as models have grown capable enough to act as useful research partners. The practice likely predates the label, and many researchers may already be doing some version of it without naming it explicitly.

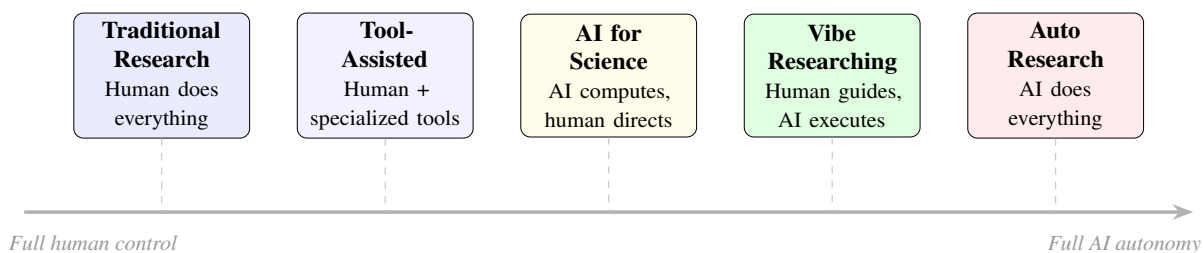
\begin{figure}[t]
\centering
\resizebox{\textwidth}{!}{%
\begin{tikzpicture}[
    every node/.style={font=\small},
    box/.style={draw, rounded corners, minimum width=2.7cm, minimum height=2.7cm, align=center, text width=2.5cm, inner sep=3pt},
]
\draw[-{Stealth[length=3mm]}, line width=1.5pt, black!50] (-8.7,-3.1) -- (8.7,-3.1);
\node[below, font=\normalsize\itshape, black!60] at (-7.3,-3.4) {Full human control};
\node[below, font=\normalsize\itshape, black!60] at (7.3,-3.4) {Full AI autonomy};

\node[box, fill=blue!8] (trad) at (-6.6,-1) {\textbf{Traditional Research}\\[1pt]{\scriptsize Human does everything}\\[2pt]\fontsize{6pt}{7pt}\selectfont\itshape no AI system; pen-and-paper, manual code};
\node[box, fill=blue!5] (tool) at (-3.3,-1) {\textbf{Tool-Assisted}\\[1pt]{\scriptsize Human + specialized tools}\\[2pt]\fontsize{6pt}{7pt}\selectfont\itshape Grammarly, Overleaf assistants, statistical software};
\node[box, fill=yellow!10] (ai4s) at (0,-1) {\textbf{AI for Science}\\[1pt]{\scriptsize AI computes, human directs}\\[2pt]\fontsize{6pt}{7pt}\selectfont\itshape AlphaFold, neural PDE solvers, AI-aided microscopy};
\node[box, fill=green!12] (vibe) at (3.3,-1) {\textbf{Vibe Researching}\\[1pt]{\scriptsize Human guides, AI executes}\\[2pt]\fontsize{6pt}{7pt}\selectfont\itshape multi-agent literature \& coding stacks (this paper)};
\node[box, fill=red!8] (auto) at (6.6,-1) {\textbf{Auto Research}\\[1pt]{\scriptsize AI does everything}\\[2pt]\fontsize{6pt}{7pt}\selectfont\itshape The AI Scientist, Agent Laboratory};

\draw[dashed, black!40] (trad.south) -- (-6.6,-3.1);
\draw[dashed, black!40] (tool.south) -- (-3.3,-3.1);
\draw[dashed, black!40] (ai4s.south) -- (0,-3.1);
\draw[dashed, black!40] (vibe.south) -- (3.3,-3.1);
\draw[dashed, black!40] (auto.south) -- (6.6,-3.1);

\end{tikzpicture}%
}
\caption{The spectrum of human-AI collaboration in research. From left to right, AI involvement increases: traditional research is fully manual; tool-assisted adds computational aids; AI for Science uses AI models for domain computation while keeping the process human-driven; vibe researching delegates the research process itself to agents; auto research automates everything.}
\label{fig:spectrum}
\end{figure}

What makes vibe researching worth examining carefully, rather than dismissing it as just ``using ChatGPT for research,'' is the set of deeper questions it raises. It sits at a specific point on the spectrum of human-AI collaboration (Fig.~\ref{fig:spectrum}), distinct from both traditional research and fully autonomous systems, and understanding where it sits matters for how we think about credit, training, quality control, and the future of scientific work. It is also distinct from ``AI for Science,'' which uses AI models for domain-specific computation without necessarily changing the research process itself (Section~\ref{sec:distinction_ai4s}). In the perspective advanced here, vibe researching is noteworthy because it reorganizes the process itself.

This article is offered as a forward-looking visionary piece on an emerging research paradigm. The contribution we aim for is to characterize and frame the practice clearly enough that the community can discuss it from shared ground. This perspective article tries to pin down what vibe researching is and why the term may be analytically useful. We are not claiming to have invented the concept or to provide a definitive taxonomy. Instead, after surveying the relevant landscape (Section~\ref{sec:related_work}), we aim to (1) give vibe researching a clear working definition, an explicit boundary criterion separating it from traditional tool-assistance, and a set of principles, and distinguish it from both AI for Science and auto research (Section~\ref{sec:definition}); (2) outline a conceptual framework for how it works, from multi-agent architectures and memory systems to the human's role in the loop, including a concrete reference architecture for vibe-researching systems (Section~\ref{sec:methodology}); (3) examine eight technical limitations, including a new dedicated discussion of prompt-injection and adversarial retrieved content (Section~\ref{sec:limitations}); (4) weigh the positive and negative societal impacts, and propose concrete reconciliation strategies for the most consequential tensions among them (Section~\ref{sec:impact}); and (5) map each limitation and negative impact to a concrete future direction, including a controlled-empirical-evaluation challenge and a dedicated agent-security research agenda (Section~\ref{sec:future}).

For readers who reach this figure expecting a sharper contrast between AI for Science and vibe researching: the two paradigms differ in \emph{where AI is inserted}. AI-for-Science systems (e.g., AlphaFold~\cite{jumper2021alphafold}, neural PDE solvers~\cite{li2020fourier}) insert a domain-specific machine-learning model into a single computational step of an otherwise traditional research process; vibe researching, by contrast, inserts a general-purpose language-and-tool agent into the \emph{process itself}, restructuring how literature is surveyed, code is written, data is analyzed, and manuscripts are drafted, while leaving domain models unchanged. Table~\ref{tab:comparison} makes the corresponding division of labor explicit, and Section~\ref{sec:distinction_ai4s} develops the contrast in detail.

\begin{table}[t]
\centering
\caption{Division of labor across research paradigms. AI for Science uses AI for domain computation but leaves the research process to humans. Vibe researching delegates process execution to agents. Auto research automates both.}
\label{tab:comparison}
\footnotesize
\resizebox{\textwidth}{!}{%
\begin{tabular}{@{}lccccc@{}}
\toprule
\textbf{Dimension} & \textbf{Traditional} & \textbf{Tool-Assisted} & \textbf{AI for Science} & \textbf{Vibe Researching} & \textbf{Auto Research} \\
\midrule
Idea generation & Human & Human & Human & Human & AI \\
Literature review & Human & Semi-auto & Human & AI (human-guided) & AI \\
Experiment design & Human & Human & Human & Human + AI & AI \\
Domain computation & Human & Human + tools & \textbf{AI model} & AI agent + tools & AI \\
Implementation & Human & Human + tools & Human & AI (human-reviewed) & AI \\
Data analysis & Human & Human + tools & Human + AI & AI (human-directed) & AI \\
Writing & Human & Human & Human & AI (human-edited) & AI \\
Quality control & Human & Human & Human & Human & AI (or none) \\
Accountability & Human & Human & Human & Human & Unclear \\
\bottomrule
\end{tabular}%
}
\end{table}

\section{Related Work}
\label{sec:related_work}

\subsection{Foundation: Large Language Models}
\label{sec:llm_foundations}

Vibe researching is made possible by a specific set of capabilities that emerged in large language models over the past several years. Early LLMs like GPT-3 \citep{brown2020gpt3} demonstrated that scaling up language models produced surprisingly general abilities, but they were difficult to steer and prone to incoherent outputs. The introduction of instruction tuning and reinforcement learning from human feedback (RLHF) \citep{ouyang2022instructgpt} changed this: models like ChatGPT and GPT-4 \citep{openai2023gpt4} became reliably responsive to natural-language instructions, making them usable as interactive tools rather than mere text generators.

Several capabilities are particularly relevant for research applications. \textit{Instruction following} allows researchers to direct agents through natural language rather than formal specifications. \textit{Long-context processing} allows models to ingest and reason over entire papers or codebases in a single session (Claude supports up to one million tokens \citep{anthropic2024claude}; Gemini \citep{gemini2024} offers similar scales). \textit{Code generation} \citep{chen2021codex} lets agents write, debug, and execute programs rather than just discuss them. \textit{Tool use}, the ability to call external APIs as part of a reasoning chain \citep{schick2023toolformer}, connects models to the outside world: search engines, databases, interpreters. \textit{Multi-turn coherence} enables the sustained back-and-forth conversations that vibe researching depends on. And \textit{chain-of-thought reasoning} \citep{wei2022chainofthought} lets models decompose complex requests into step-by-step plans.

Taken together, these capabilities did not become broadly usable in their current form until the early 2020s. Their convergence in current frontier models is what makes the practice described in this perspective plausible, and their continued improvement will shape how far it can go.

\subsection{Vibe Coding}
\label{sec:vibe_coding}

The term ``vibe coding'' was coined by Andrej Karpathy in February 2025 \citep{karpathy2025vibecoding}. He described a workflow built on LLM coding assistants like Cursor with Claude Sonnet: the developer never opens a diff, error messages get copy-pasted back without commentary, and the ``Accept All'' button becomes the primary interaction. Karpathy was explicit that this was meant for throwaway prototypes, but the term resonated far beyond that context because it named a shift many developers were already experiencing: from \textit{writing} code to \textit{directing} code generation.

The conceptual parallel to research is striking. Just as vibe coding delegates code production to an LLM while the developer steers by intent and evaluation, vibe researching delegates research execution (literature search, implementation, data analysis, drafting) to LLM agents while the researcher steers by domain judgment and critical assessment. The key difference is stakes: a buggy prototype costs an afternoon, while a flawed research claim can mislead an entire field. This is why vibe researching cannot simply inherit vibe coding's ``accept all'' ethos; it must add the verification and oversight layers described in Section~\ref{sec:methodology}. Nevertheless, the underlying cognitive shift is the same: from producing artifacts to directing their production and evaluating their quality.

\subsection{AI in the Research Pipeline}
\label{sec:ai_research_pipeline}

The application of LLMs to scientific research has progressed rapidly along a spectrum from narrow tools to broad autonomous systems. Ren et al.~\cite{ren2025scientificintelligence} survey this landscape, tracing the arc from simple prompt-and-answer usage to multi-step agents that can plan and execute across entire research pipelines.

At the tool end of the spectrum, systems target specific bottlenecks. LitLLM \citep{agarwal2024litllm} automates literature review through keyword and embedding-based search with LLM re-ranking. Paper Copilot \citep{lin2024papercopilot} builds personalized research profiles and tracks trending topics. Benchmarks such as MLAgentBench \citep{huang2023mlagentbench}, MLE-bench \citep{chan2024mlebench}, and AgentBench \citep{liu2023agentbench} measure how well LLM agents handle research-adjacent tasks. Two further entries round out the tool end of the spectrum: STORM~\cite{shao2024storm}, which automates multi-perspective, Wikipedia-style literature synthesis at the writing stage, and LAB-Bench~\cite{laurent2024labbench}, a research-grade benchmark that measures agent capabilities on real biology-research tasks rather than proxy ones.

Further along the spectrum, domain-specific agents have produced real experimental results. ChemCrow \citep{bran2024chemcrow} wraps an LLM with 18 chemistry tools and has autonomously planned and executed chemical syntheses, including a novel chromophore. Coscientist \citep{boiko2023coscientist}, published in \textit{Nature}, showed GPT-4 designing and running palladium-catalyzed cross-coupling reactions with minimal human input. SciAgents \citep{ghafarollahi2024sciagents} combines multi-agent architectures with ontological knowledge graphs to automate discovery in materials science.

At the far end sits fully autonomous research. The AI Scientist \citep{lu2024aiscientist} generates ideas, writes and runs code, produces figures, drafts papers, and simulates peer review, all for under \$15 per paper. Its successor, The AI Scientist-v2 \citep{yamada2025aiscientistv2}, used agentic tree search and was reported to have produced an entirely AI-generated paper accepted at an ICLR workshop. Agent Laboratory \citep{schmidgall2025agentlab} accepts a human-provided idea and autonomously handles the rest, reporting an 84\% cost reduction. MLR-Copilot \citep{li2024mlrcopilot} follows a similar pipeline for machine learning research. These results are notable, but Agent Laboratory's own evaluation \citep{schmidgall2025agentlab} also notes that a substantial fraction of AI Scientist experiments fail due to coding errors, and the quality of the surviving papers remains limited.

Table~\ref{tab:systems} summarizes representative systems across this spectrum. Vibe researching draws on all three levels. It uses narrow tools for specific tasks, leverages agent capabilities for multi-step execution, but stops short of full autonomy by keeping the human in the orchestration role.

A closely related prior work that uses the same terminology is Zhang~\cite{wolfcoming}, who introduces ``vibe researching'' in the context of social-science research and proposes a task-typology framework structured around \emph{codifiability} (the extent to which a research task can be reduced to explicit, executable steps) and \emph{tacit knowledge} (the extent to which it depends on expert judgment that resists codification). That work is domain-restricted to social-science research and task-typology focused: for each kind of social-science task, it asks whether AI agents are positioned to augment or replace the human researcher, and concludes that agents excel at speed, coverage, and methodological scaffolding but struggle with theoretical originality and tacit field knowledge. The present paper takes a complementary cross-domain perspective. Rather than asking which tasks within a single discipline are codifiable, we ask how the \emph{research process itself}, shared across disciplines, is restructured when LLM-based agents are inserted into it, what \emph{technical mechanisms} enable that restructuring (Section~\ref{sec:methodology}), what its \emph{limitations} are at the agent and infrastructure level (Section~\ref{sec:limitations}), and what \emph{societal and educational consequences} follow (Section~\ref{sec:impact}). The two analyses are complementary: their codifiability-vs-tacit-knowledge axis is a natural lens for deciding, within any given discipline, where on our human-agent spectrum (Fig.~\ref{fig:spectrum}) a particular research task should sit.

\begin{table}[t]
\centering
\caption{Representative AI-assisted research systems across the automation spectrum, from single-task tools to fully autonomous pipelines.}
\label{tab:systems}
\footnotesize
\resizebox{\textwidth}{!}{%
\begin{tabular}{@{}p{2.6cm}p{1.9cm}p{1.7cm}p{1.3cm}p{3.4cm}@{}}
\toprule
\textbf{System} & \textbf{Scope} & \textbf{Human Role} & \textbf{Domain} & \textbf{Key Capability} \\
\midrule
LitLLM \citep{agarwal2024litllm} & Literature review & Directs search & General & RAG with LLM re-ranking \\
Paper Copilot \citep{lin2024papercopilot} & Literature tracking & Sets profile & General & Personalized paper recommendation \\
STORM \citep{shao2024storm} & Writing/\allowbreak{}synthesis & Provides topic & General & Multi-perspective Wikipedia-style synthesis \\
ChemCrow \citep{bran2024chemcrow} & Experiment execution & Minimal & Chemistry & 18 integrated chemistry tools \\
Coscientist \citep{boiko2023coscientist} & Experiment execution & Minimal & Chemistry & Autonomous wet-lab reactions \\
SciAgents \citep{ghafarollahi2024sciagents} & Discovery & Minimal & Materials & Knowledge graph reasoning \\
AI Scientist \citep{lu2024aiscientist} & End-to-end & None & ML & Full pipeline for \$15/paper \\
Agent Laboratory \citep{schmidgall2025agentlab} & End-to-end & Provides idea & ML & 84\% cost reduction \\
AutoGen / MetaGPT / CAMEL \citep{wu2023autogen,hong2023metagpt,li2023camel} & Agent framework & Designs roles & General & Multi-agent orchestration substrate \\
LAB-Bench \citep{laurent2024labbench} & Benchmark & Evaluator & Biology & Research-grade capability evaluation \\
\midrule
\textbf{Vibe Researching} & \textbf{Process-wide} & \textbf{Orchestrates} & \textbf{General} & \textbf{Human as meta-agent} \\
\bottomrule
\end{tabular}%
}
\end{table}

\subsection{Agent Frameworks and Infrastructure}

Alongside research-specific systems, a parallel line of work has developed general-purpose frameworks for building and coordinating LLM agents. AutoGen \citep{wu2023autogen} enables multi-agent conversations where specialized agents collaborate through structured protocols. MetaGPT \citep{hong2023metagpt} assigns distinct roles (engineer, product manager, architect) to separate agents and coordinates them through standardized outputs. CAMEL \citep{li2023camel} explores how pairs of agents can collaborate through role-playing to solve complex tasks. These frameworks provide the coordination infrastructure that vibe researching depends on.

Equally important are advances in agent memory and grounding. MemGPT \citep{packer2023memgpt} applies operating-system-style virtual memory management to extend an agent's effective context beyond its native window. The Generative Agents architecture \citep{park2023generativeagents} demonstrates how agents can synthesize observations into higher-level reflections and retrieve them when relevant. Voyager \citep{wang2023voyager} introduces a skill library where agents store verified procedures for reuse. These memory and skill mechanisms are essential for research projects that span days to months, as discussed in Section~\ref{sec:methodology}.

\subsection{Human-AI Collaboration}

The question of how humans and AI systems should work together predates LLMs. Parasuraman et al.~\cite{parasuraman2000automation} proposed an influential taxonomy of automation levels, ranging from fully manual to fully autonomous, with intermediate stages where the system suggests, recommends, or executes with varying degrees of human oversight. This framework remains useful for understanding where vibe researching sits: it occupies a middle level where the agent executes tasks but the human retains decision authority and can override at any point. Lee and See~\cite{lee2004trust} established that appropriate trust calibration is essential for effective human-automation interaction; overtrust leads to uncritical acceptance of errors, while undertrust leads to disuse. Both failure modes are directly relevant to vibe researching (see Section~\ref{sec:verification_asymmetry}).

With the arrival of foundation models, these questions have become newly urgent. Vats et al.~\cite{vats2024humanai} survey human-AI collaboration with foundation models and reach a sobering conclusion: effective collaboration does not fall out automatically from bigger models. It requires deliberate, human-centered design, with particular attention to trust, interpretability, and alignment between what the AI produces and what the human needs. Bansal et al.~\cite{bansal2021complementary} show experimentally that human-AI teams do not always outperform either alone; complementary performance requires that the human can accurately judge when to rely on the AI and when to override it. Noy and Zhang~\cite{noy2023productivity} provide experimental evidence that generative AI substantially increases productivity in writing tasks, with the largest gains for lower-skilled workers, suggesting that AI assistance may compress the skill distribution rather than uniformly raising it.

A useful distinction from this literature is between ``human-in-the-loop'' systems (where the AI drives and the human supervises) and ``AI-in-the-loop'' systems (where the human drives and the AI assists). The fully autonomous systems described in Section~\ref{sec:ai_research_pipeline} go further still, minimizing or eliminating the human role entirely. Vibe researching falls squarely in the AI-in-the-loop camp: the researcher sets the agenda and the agents support it, not the other way around. This embedded oversight is what distinguishes vibe researching from automation and is essential for maintaining the intellectual accountability that scientific work demands, as we detail in Section~\ref{sec:methodology}.

\section{What Is Vibe Researching?}
\label{sec:definition}

\subsection{A Working Definition}

\begin{quote}
\textbf{Vibe researching} is a mode of scientific inquiry in which a human researcher provides high-level direction, creative intuition, and critical evaluation, while LLM-based agents carry out the labor-intensive parts of the process (literature discovery, experimental implementation, data analysis, manuscript preparation) in response to natural language instructions.
\end{quote}

The word ``vibe'' is doing real work here. It points to the quality of the human's engagement: not line-by-line control, but a sustained \textit{feel} for where the research is heading. The researcher intervenes to steer, evaluate, and course-correct, contributing taste, judgment, and vision. The agent contributes speed, breadth, and tireless execution.

In these terms, the qualifier ``vibe'' names a working mode of \emph{macro-level} guidance rather than \emph{micro-level} control. The researcher steers by problem choice, evaluative judgment, and course correction informed by domain intuition and holistic understanding, while the intermediate steps (which prompt to issue, which line of code to write, how to phrase a given paragraph) fall to the agent. A researcher who instead reads, edits, and rewrites every prompt and every output sentence by sentence is exercising micro-level control: the agent has become a typing aid, and the person is still doing the underlying cognitive work directly, much as in traditional tool-assisted research. Both modes are legitimate and both are common in practice; naming the macro-level one matters because the technical, social, and educational consequences examined in the rest of this paper follow from delegating the intermediate work rather than supervising each step. Section~\ref{sec:boundary} makes the resulting boundary with tool-assisted research precise.

In terms of the automation taxonomy proposed by Parasuraman et al.~\cite{parasuraman2000automation}, vibe researching operates at an intermediate level: the agent generates options and executes selected plans, but the human retains authority over what gets executed and whether the result is acceptable. This places it above simple tool-assistance (where the human does the work and the tool helps) but below full automation (where the system acts independently). The distinction matters because the appropriate level of automation depends on the cost of errors \cite{lee2004trust}, and in scientific research, errors can propagate through the literature with lasting consequences.

\subsection{Core Principles}

Five principles, drawn from observing how the practice actually works, help distinguish vibe researching from neighboring paradigms:

\begin{enumerate}
    \item \textbf{Human as Creative Director.} The researcher chooses the questions, judges what findings matter, and makes the strategic calls. The higher-order cognitive work (recognizing novelty, connecting disparate ideas, assessing importance) stays with the person.

    \item \textbf{Natural Language as the Primary Interface.} The researcher talks to agents in plain language: ``survey the last two years of work on X,'' ``run this ablation,'' ``draft a related-work section covering Y and Z.'' The barrier to entry is a sentence, not a script. That said, learning to frame requests well is itself a skill (see Section~\ref{sec:methodology}), but the medium remains natural language rather than formal specifications.

    \item \textbf{Delegation with Oversight.} Tasks get handed off to agents, but the researcher reviews what comes back. This is where vibe researching parts company with vibe coding: Karpathy's ``accept all'' ethos does not transfer to science. You can ship a buggy prototype; you cannot publish a paper you have not actually read.

    \item \textbf{Iterative Refinement.} The work proceeds in loops. The researcher reads agent output, pushes back (``this misses the key comparison,'' ``try a different angle''), and the agent revises. Each cycle tightens the focus.

    \item \textbf{Human Accountability.} At the end of the day, the researcher's name goes on the paper and the researcher answers for every claim. AI agents are instruments, not co-authors, and the human must be able to defend every conclusion in the work. This aligns with emerging ethical guidelines that place responsibility for AI-assisted output squarely on the human author \citep{porsdammann2024ethical}.
\end{enumerate}

\subsection{Boundary with Tool-Assisted Research}
\label{sec:boundary}

The five principles above also describe, in spirit, many interactive AI assistance tools that have existed for some time, so it is fair to ask where ordinary tool-assistance ends and vibe researching begins. Without a clear boundary, the term becomes diffuse and the analytical work of this paper loses purchase. We therefore propose an explicit three-part criterion. An activity belongs to vibe researching when all three conditions hold:

\begin{enumerate}[noitemsep,label=(\roman*)]
    \item \textbf{Multi-step, goal-directed delegation.} The researcher hands off a task that requires several actions to complete (planning, retrieval, computation, synthesis, or some combination) rather than a single-turn lookup or transformation. ``Search arXiv for related work on $X$, summarize the five most relevant papers, and identify any methodological consensus'' is multi-step; ``rephrase this sentence'' is not.
    \item \textbf{Autonomy between human checkpoints.} The agent acts autonomously across the intermediate steps, using planning, memory, and tool use to make local decisions. The human inspects checkpoints (after retrieval, after implementation, after analysis, after drafting) rather than approving each micro-step. If the human is in the loop on every individual operation, the agent is being used as a typing aid and the activity sits on the tool-assistance side of the line.
    \item \textbf{Natural language as primary specification.} The instruction is given in natural language and remains the canonical statement of intent; it is the agent (not the human) that decides the concrete operational steps. Activities that require the human to author a formal program, script, or pipeline before the agent runs are tool-assisted at heart, even if a language model helps the human write that artifact.
\end{enumerate}

Activities that fail any of the three conditions belong to traditional tool-assistance. Two worked examples make the line concrete. \emph{Example~A:} a researcher asks ChatGPT to rephrase a paragraph for clarity and copies the result back into the manuscript. This is a single-turn transformation with no autonomous intermediate steps; it fails condition~(i) and is tool-assistance, not vibe researching. \emph{Example~B:} a researcher instructs a multi-agent stack to ``survey the recent literature on parameter-efficient fine-tuning, identify the three most promising approaches, reimplement them on our internal benchmark, and produce a comparison table.'' The agents plan the search, retrieve papers, write and debug code, run experiments, and assemble the output across hours of autonomous work; the researcher inspects the checkpoints and redirects as needed. This satisfies all three conditions and is paradigmatic vibe researching.

A continuum exists between these two extremes (for example, asking an agent to draft a related-work section with citations, where the agent autonomously retrieves and integrates sources but the task is bounded). The criterion is intended to mark a threshold, not to claim that intermediate activities do not exist. Where an activity sits on the continuum determines how many of this paper's technical limitations and social consequences apply to it; activities further toward the vibe-researching end inherit more of them.

\subsection{Distinction from AI for Science}
\label{sec:distinction_ai4s}

It is worth clarifying how vibe researching relates to the broader ``AI for Science'' movement \citep{zhang2024scientificllm}. AI for Science typically refers to using machine learning models as domain-specific computational tools within a research project: AlphaFold for protein structure prediction \citep{jumper2021alphafold}, graph neural networks for molecular property estimation \citep{gilmer2017mpnn}, neural PDE solvers for accelerating simulations \citep{li2020fourier}. In these applications, AI solves a \textit{specific technical problem} embedded in the research, but the research process itself remains unchanged. The scientist still reads the literature, designs the experiments, interprets the results, and writes the paper by hand.

The word ``tool'' can blur the boundary, so the distinction is worth making explicit. In our usage, tool-assisted research names an interaction pattern: the human researcher still executes and supervises the research process step by step, using software to make particular operations easier or faster. AI for Science names a computational role: a learned model performs a domain-specific scientific computation, such as predicting a protein structure, solving a PDE, or analyzing microscopy data. Thus, an AI-for-Science system may be used as a tool in the everyday sense, but it differs from ordinary tool-assistance because the AI model substitutes for the domain computation rather than merely helping the human operate the workflow. Vibe researching shifts the boundary again: the delegated object is not only one domain computation but the surrounding research process itself, including literature work, implementation, analysis, and drafting.

Vibe researching operates at a different level. Rather than replacing one computational step, it restructures the entire workflow by inserting LLM agents into the research \textit{process}: literature review, experimental design, implementation, analysis, and writing. The AI is not a domain-specific model trained on protein sequences or molecular graphs; it is a general-purpose language agent directed through natural language conversation.

The two are complementary, not competing. A vibe-researching session might involve directing an agent to set up and run AlphaFold, analyze the outputs, and draft a results section. The AI-for-Science model (AlphaFold) handles the domain computation; the vibe-researching agent handles everything around it. Recognizing this distinction matters because it clarifies what vibe researching is \textit{not}: it is not a new way to do computation, but a new way to organize and execute the research process.

\subsection{Comparison with Other Paradigms}

To see where vibe researching sits relative to other research paradigms, it helps to look at the landscape from two angles: who does what (Table~\ref{tab:comparison}), and how the system is wired (Table~\ref{tab:method_compare}).

Table~\ref{tab:comparison} shows the division of labor. The AI-for-Science column is instructive: AI handles domain-specific computation (the bolded cell) but the rest of the process remains manual. Vibe researching delegates the process itself to agents while keeping quality control and accountability with the human. Auto research hands over both.

\begin{table}[t]
\centering
\caption{Architectural differences between vibe researching and auto research. Both use the same underlying techniques; the distinction is who orchestrates them.}
\label{tab:method_compare}
\footnotesize
\resizebox{\textwidth}{!}{%
\begin{tabular}{@{}lp{4.8cm}p{4.8cm}@{}}
\toprule
\textbf{Aspect} & \textbf{Vibe Researching} & \textbf{Auto Research} \\
\midrule
Orchestration & Human decides what to do next & Meta-agent or fixed pipeline controls flow \\[3pt]
Goal specification & Open-ended, evolves through conversation & Defined upfront as a structured input \\[3pt]
Error recovery & Agent self-checks + human redirects & Agent self-corrects or fails silently \\[3pt]
Quality control & Human checkpoints, aided by agent verification & Automated evaluation (e.g., simulated review) \\[3pt]
Scope of autonomy & Per-task; human grants and revokes & End-to-end; runs to completion \\[3pt]
Adaptability & High; researcher can pivot at any moment & Low; pipeline structure is fixed \\[3pt]
Reproducibility & Requires interaction logging & Fixed pipeline is more easily reproducible \\
\bottomrule
\end{tabular}%
}
\end{table}

Table~\ref{tab:method_compare} goes deeper. Vibe researching and auto research share the same technical substrate: multi-agent architectures, memory, tool use, planning, retrieval. The difference is not technological but architectural, and it centers on who plays the role of orchestrator.

In auto-research systems like The AI Scientist \citep{lu2024aiscientist} and the framework proposed by Liu et al.~\cite{liu2025autoresearch}, a meta-agent or a pre-defined pipeline coordinates the specialist agents. The system receives a high-level goal and runs autonomously until it produces a paper. The human, if involved at all, appears only at the end to review the output. Vibe researching inverts this: the human \textit{is} the meta-agent. They observe the output of each step, decide whether to proceed or backtrack, and adjust the direction in real time.

Empirical evidence from other knowledge-work domains supports this design choice: Noy and Zhang~\cite{noy2023productivity} found that generative AI significantly boosts productivity but that the quality of output depends on how actively the human engages with the AI's contributions rather than accepting them passively. This makes vibe researching less efficient in wall-clock terms (the human is a bottleneck) but more robust to the kinds of errors that compound in autonomous pipelines: a flawed experimental setup that goes unnoticed, a misframed question that leads to irrelevant experiments, or a subtle statistical error that a simulated reviewer does not catch. The trade-off is between throughput and reliability. Auto research optimizes for the former; vibe researching bets that the latter matters more in science, at least for now.

\section{Conceptual Framework}
\label{sec:methodology}

This section outlines a conceptual framework for vibe researching rather than a validated method in the experimental sense. We begin with the overall interaction structure, then describe the agent-side techniques that make the pattern plausible, and finally examine the human's role in the loop. The discussion draws on recent work in LLM agent architectures \citep{sumers2023coala, wang2024survey} and systems designed for automated research \citep{liu2025autoresearch, schmidgall2025agentlab}.

\subsection{Framework: The Human-Agent Interaction Loop}

At its core, vibe researching can be described as a simple loop. The researcher thinks, the agent acts, and the researcher decides what happens next. More precisely, each interaction cycle can be described in five steps:

\begin{enumerate}[noitemsep]
    \item \textbf{Instruct.} The researcher issues a natural-language directive: a question to investigate, a task to carry out, or a critique of prior output.
    \item \textbf{Execute.} The agent (or a team of agents) plans and carries out the task, invoking tools, consulting memory, and producing intermediate artifacts.
    \item \textbf{Present.} The agent returns results: text, code, data, figures, or a summary of actions taken.
    \item \textbf{Evaluate.} The researcher reviews the output for correctness, relevance, and quality.
    \item \textbf{Redirect.} Based on the evaluation, the researcher either accepts the output and moves on, requests revisions, or changes direction entirely.
\end{enumerate}

This loop is deliberately asymmetric. The researcher spends most of their effort on steps 1, 4, and 5 (the cognitive work), while the agent bears much of the burden of steps 2 and 3 (the execution). In a productive session, a researcher may move through many such cycles in quick succession, exploring the problem space faster than would usually be practical without agent support.

At a coarser granularity, these micro-cycles often organize into a five-phase workflow (Fig.~\ref{fig:framework}):

\begin{figure}[t]
\centering
\resizebox{\textwidth}{!}{%
\begin{tikzpicture}[
    node distance=0.8cm and 0.5cm,
    phase/.style={draw, rounded corners, fill=blue!6, minimum width=2.4cm, minimum height=1cm, align=center, font=\small},
    arrow/.style={-{Stealth[length=2.5mm]}, thick},
    feedback/.style={-{Stealth[length=2.5mm]}, thick, dashed, red!60!black},
]
\node[phase] (p1) {\textbf{Phase 1}\\Ideation};
\node[phase, right=0.8cm of p1] (p2) {\textbf{Phase 2}\\Exploration};
\node[phase, right=0.8cm of p2] (p3) {\textbf{Phase 3}\\Experimentation};
\node[phase, right=0.8cm of p3] (p4) {\textbf{Phase 4}\\Synthesis};
\node[phase, right=0.8cm of p4] (p5) {\textbf{Phase 5}\\Refinement};
\draw[arrow] (p1) -- (p2);
\draw[arrow] (p2) -- (p3);
\draw[arrow] (p3) -- (p4);
\draw[arrow] (p4) -- (p5);
\draw[feedback] (p3.south) -- ++(0,-0.7) -| (p2.south);
\draw[feedback] (p5.south) -- ++(0,-0.7) -| (p3.south);
\node[above=0.15cm of p1, font=\scriptsize\itshape, blue!70!black] {Human-led};
\node[above=0.15cm of p2, font=\scriptsize\itshape, green!50!black] {AI-executed};
\node[above=0.15cm of p3, font=\scriptsize\itshape, green!50!black] {AI-executed};
\node[above=0.15cm of p4, font=\scriptsize\itshape, green!50!black] {AI-executed};
\node[above=0.15cm of p5, font=\scriptsize\itshape, blue!70!black] {Human-led};
\coordinate (mid) at ($(p2.south)!0.5!(p5.south)$);
\node[font=\scriptsize, below] at (mid |- 0,-1.5) {Dashed arrows indicate iterative feedback loops};
\end{tikzpicture}%
}
\caption{A typical five-phase vibe-researching workflow. Blue phases are human-led; green phases are AI-executed with human oversight. Dashed arrows show common feedback loops.}
\label{fig:framework}
\end{figure}
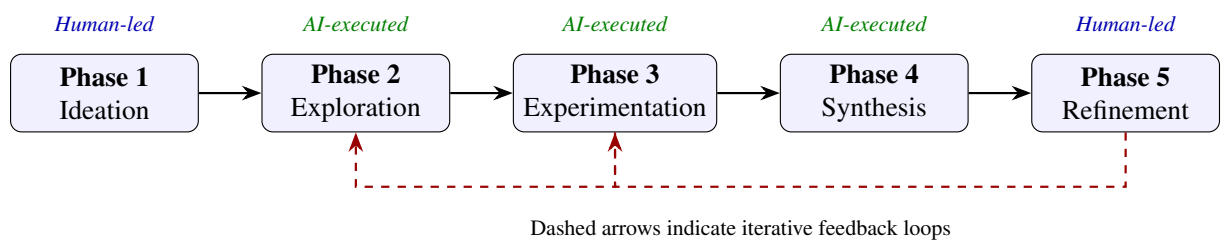

\begin{itemize}[noitemsep]
    \item \textbf{Ideation} (human-led): the researcher picks the research question.
    \item \textbf{Exploration} (AI-executed): agents survey the literature and map the landscape.
    \item \textbf{Experimentation} (AI-executed): agents implement and run experiments under the researcher's direction.
    \item \textbf{Synthesis} (AI-executed): agents compile findings into draft text, tables, and figures.
    \item \textbf{Refinement} (human-led): the researcher takes ownership of the final output, verifying every claim.
\end{itemize}

Not every project follows these phases in strict sequence; the dashed arrows in the figure reflect the reality that researchers frequently loop back. The critical invariant is that the human controls the entry and exit of each phase, while agents handle the work within it.

\subsection{Enabling Techniques}
\label{sec:enabling_techniques}

The framework above defines the rhythm of vibe researching; the techniques below help make it technically feasible in practice. All of them build on the foundational LLM capabilities discussed in Section~\ref{sec:llm_foundations} (instruction following, code generation, tool use, long-context processing, multi-turn coherence, and chain-of-thought reasoning), which together make it possible to treat an LLM as a research assistant rather than only a search engine. The techniques below address specific challenges in the research workflow that these raw capabilities alone do not solve. Table~\ref{tab:techniques} provides an overview.

\begin{table}[t]
\centering
\caption{Enabling techniques for vibe-researching agents and their roles in the research workflow.}
\label{tab:techniques}
\footnotesize
\resizebox{\textwidth}{!}{%
\begin{tabular}{@{}p{2.5cm}p{5.8cm}p{4.5cm}@{}}
\toprule
\textbf{Technique} & \textbf{Why It Matters for Research} & \textbf{Representative Work} \\
\midrule
Multi-agent systems & Specialization across tasks; distributes context to avoid window limits; enables parallel execution and structured coordination & AutoGen \citep{wu2023autogen}, MetaGPT \citep{hong2023metagpt}, CAMEL \citep{li2023camel} \\ \hline
Memory mechanisms & Projects span days to months; agents must remember prior decisions, accumulated findings, and conventions across sessions & MemGPT \citep{packer2023memgpt}, Generative Agents \citep{park2023generativeagents} \\ \hline
Tool use \& skills & Agents must query APIs, execute code, process data, compile \LaTeX{}, and build reusable procedures for recurring tasks & Toolformer \citep{schick2023toolformer}, ToolLLM \citep{qin2023toolllm}, Voyager \citep{wang2023voyager} \\ \hline
Planning \& decomposition & Complex requests require breaking into ordered subtasks, executing each, and adapting when steps fail & ReAct \citep{yao2022react}, CoT \citep{wei2022chainofthought} \\ \hline
RAG & Grounds output in real papers and data; enables literature search, fact-checking, and evidence-based writing & LitLLM \citep{agarwal2024litllm} \\ \hline
Self-reflection & Agents check their own code for errors, flag inconsistencies with prior results, and signal uncertainty to the researcher & Reflexion \citep{shinn2023reflexion} \\
\bottomrule
\end{tabular}%
}
\end{table}

\subsubsection{Multi-Agent Architecture}
\label{sec:multi_agent}

Research tasks naturally decompose into roles with different capabilities. Searching literature is not the same skill as debugging code or drafting a related-work section. A single monolithic agent asked to do all of these tends to lose focus, forget context, and make more errors than a team of specialists. Splitting work across multiple agents also helps manage context window constraints (Section~\ref{sec:limitations}): each agent carries only the context relevant to its role rather than the entire project state. Multiple agents can work in parallel (e.g., one surveying literature while another sets up an experimental pipeline), and the division of labor provides natural structure for planning and coordination.

Multi-agent frameworks such as AutoGen \citep{wu2023autogen}, MetaGPT \citep{hong2023metagpt}, and CAMEL \citep{li2023camel} have shown that assigning distinct roles to separate agents and letting them communicate through structured protocols produces better results than a generalist approach. In a vibe-researching setup, the natural division looks something like:

\begin{itemize}[noitemsep]
    \item A \textit{literature agent} that searches academic databases, retrieves papers, and produces structured summaries.
    \item A \textit{coding agent} that writes, tests, and debugs experimental code.
    \item An \textit{analysis agent} that runs statistical tests, generates plots, and interprets results.
    \item A \textit{writing agent} that drafts and revises manuscript text.
\end{itemize}

Liu et al.~\cite{liu2025autoresearch} formalize a similar decomposition in their Agent-Based Auto Research framework, coordinating specialized agents across the full research lifecycle from literature review through dissemination. The key difference in vibe researching is that the human, not a meta-agent, decides when to invoke each specialist and whether the output is acceptable.

Multi-agent decomposition is not a free lunch. Talebirad and Nadiri~\cite{talebirad2023multiagent}, in their multi-agent LLM collaboration framework, explicitly enumerate looping issues, scalability constraints, and conflicts/inconsistencies among agents as central challenges of multi-agent systems. In a research context the most damaging of these is silent inter-agent inconsistency: a literature agent that retrieves papers under one definition of the topic and an analysis agent that operates under a slightly different definition can produce a coherent-looking output whose underlying premise is muddled, and the bug is invisible to either agent in isolation. Multi-agent architectures help most when the sub-tasks are clearly bounded, the inter-agent interfaces carry typed structured data rather than free-form prose, and the orchestrator (whether human or meta-agent) retains visibility into each handoff.

\subsubsection{Memory Mechanisms}
\label{sec:memory_mechanisms}

Research projects are stateful over long time horizons. A literature review builds context that informs experimental design; experimental results shape the narrative of the paper; design decisions made in week one constrain what is possible in week four. LLM agents, constrained by finite context windows, need explicit memory to maintain coherence across sessions that may span days to months.

Zhang et al.~\cite{zhang2024memorymechanism} survey memory mechanisms in LLM-based agents, identifying three layers that map naturally onto vibe researching:

\begin{enumerate}[noitemsep]
    \item \textbf{Working memory}: the agent's current context window, holding the active conversation, the file being edited, and the most recent outputs. Fast but limited.
    \item \textbf{Episodic memory}: the history of past interactions, including prior prompts, agent responses, and intermediate results. When the researcher says ``go back to the approach we tried yesterday,'' the system retrieves relevant episodes.
    \item \textbf{Semantic memory}: structured knowledge accumulated over the project: key findings, paper summaries, established conventions, and design decisions. This is the project's institutional memory, and losing it mid-project can mean weeks of rework.
\end{enumerate}

MemGPT \citep{packer2023memgpt} demonstrates how virtual memory management, inspired by operating-system paging, can extend an agent's effective context far beyond its native window. The Generative Agents architecture \citep{park2023generativeagents} shows how agents can synthesize raw observations into higher-level reflections and retrieve them when relevant, a pattern directly useful for maintaining research context over long sessions.

In practice, current vibe-researching setups often rely on simpler mechanisms: project files the agent reads at the start of each session, conversation logs that can be resumed, and explicit researcher instructions like ``here is what we established last time.'' The gap between what memory systems promise and what they deliver is one of the more frustrating aspects of current practice (see Section~\ref{sec:limitations}). As these systems mature, the handoff between sessions should become smoother, but the researcher's role in curating what the agent remembers (and forgets) remains important.

Episodic memory in particular involves a sharp trade-off between retrieval precision and context cost. A coarse retrieval policy injects many past episodes into the working context: recall is high, but the model spends scarce context on irrelevant material and can be misled by stale prior decisions that no longer apply. A precise retrieval policy injects only the few most-relevant episodes: context is preserved, but a relevant episode that misses the retrieval threshold simply does not get used. There is no general optimum; what helps most is making the trade-off observable to the researcher, so that when an agent appears to have ``forgotten'' something the researcher knows whether the issue is a retrieval miss (fixable by reformulating the prompt) or a context budget overflow (fixable by trimming the working set).

\subsubsection{Tool Use and Skills}

Raw language generation is not enough for research. An agent that can only talk about code but not run it, or discuss a paper but not fetch it, is of limited use. The tool-use paradigm, where an LLM invokes external functions as part of its reasoning chain, is what makes vibe researching practical rather than merely conversational.

Toolformer \citep{schick2023toolformer} showed that language models can learn to call APIs in a self-supervised fashion. ToolLLM \citep{qin2023toolllm} scaled this to over 16,000 real-world APIs. In the research context, agents need to interact with a broad range of external systems:

\begin{itemize}[noitemsep]
    \item \textbf{Search and retrieval}: querying Semantic Scholar, arXiv, Google Scholar, and the open web to find relevant work.
    \item \textbf{Code execution}: writing and running Python/R scripts, Jupyter notebooks, and shell commands to implement and test ideas.
    \item \textbf{Data processing}: loading datasets, cleaning data, computing statistics, and generating visualizations with libraries like pandas and matplotlib.
    \item \textbf{Document preparation}: compiling \LaTeX{}, managing BibTeX references, producing publication-ready figures and tables.
\end{itemize}

A practical caveat: many of these systems were not built with agent access in mind, and the resulting friction is a significant bottleneck (see Section~\ref{sec:limitations} for details). Where APIs exist and are well-structured, tool use works well; where they do not, the researcher may need to step in or build custom wrappers. A second, subtler failure mode is the tool call itself: when many tools with overlapping functionality are available, agents frequently select the wrong tool or fabricate plausible-but-invalid arguments. Gorilla~\cite{patil2024gorilla} shows that grounding the agent in up-to-date, retrieved API documentation substantially reduces such hallucinated calls, a concrete reminder that the quality and currency of a tool's documentation effectively functions as part of the prompt.

Beyond individual tool calls, the notion of \textit{skills} (reusable, composable procedures that agents build up over time) adds another layer. Voyager \citep{wang2023voyager} introduced a skill library where an agent stores verified code snippets and retrieves them when encountering similar tasks later. In vibe researching, this pattern is directly useful: an agent that has once figured out how to set up a specific experimental pipeline, parse a particular dataset format, or compile a project with a non-standard build system can store that procedure and reuse it in future sessions without the researcher needing to explain it again. Over time, the agent accumulates a growing repertoire of project-specific capabilities.

\subsubsection{Planning and Task Decomposition}

A request like ``evaluate our method against the baselines from [paper]'' is not a single action. The agent needs to identify the baselines, find or reimplement them, set up comparable evaluation conditions, run experiments, collate results, and handle the inevitable failures along the way (a dependency that will not install, a dataset in an unexpected format, a baseline that produces NaN on the first run). This calls for planning: breaking a high-level goal into an ordered sequence of subtasks, executing each, and adapting when steps fail rather than halting.

The ReAct framework \citep{yao2022react} interleaves reasoning traces with actions, allowing agents to think through a problem step by step while gathering information from external sources. Chain-of-thought prompting \citep{wei2022chainofthought} elicits structured reasoning that helps agents decompose complex requests into manageable pieces. Together, these techniques enable agents to handle multi-step research tasks that would otherwise require the researcher to micromanage every intermediate step.

In vibe researching, planning happens at two levels. The researcher provides the macro plan (``first survey the field, then implement, then write up''), while the agent handles micro planning within each delegated task: deciding what order to try things in, recovering from errors, and re-sequencing when an earlier assumption turns out to be wrong. The researcher can intervene at any point to reprioritize, skip steps, or redirect, but the agent is expected to manage the details of execution on its own. These planning techniques should not be over-trusted, however. Evaluated as autonomous planners on standardized planning problems, current models perform poorly: Valmeekam et al.~\cite{valmeekam2023planning} report that even GPT-4 generates executable plans only around 12\% of the time. The practical implication for vibe researching is that an agent's micro-plans are most reliable over short horizons punctuated by human checkpoints, whereas long, unverified plan-and-execute chains are a characteristic failure mode that argues for human re-planning at milestones rather than a single plan fixed in advance.

\subsubsection{Retrieval-Augmented Generation}

One of the most common failure modes of LLMs in research is confident fabrication: generating plausible-sounding citations that do not exist, or stating ``facts'' not grounded in any source. Retrieval-augmented generation (RAG) addresses this by having the agent fetch relevant documents before generating a response, so that the output is grounded in actual sources rather than parametric memory alone.

In vibe researching, RAG is pervasive. When an agent surveys the literature, it does not recite papers from memory; it searches a database, retrieves real results, and summarizes them. When it writes a related-work section, it pulls from the papers it has actually read. Tools like LitLLM \citep{agarwal2024litllm} build RAG into the literature review pipeline, combining keyword search, embedding-based retrieval, and LLM re-ranking to surface relevant work.

The broader principle is that a vibe-researching agent should never be trusted to \textit{know} something; it should be equipped to \textit{look it up}. This shifts the failure mode from hallucination to retrieval error, which is somewhat easier to detect but by no means eliminated; Section~\ref{sec:limitations} discusses the residual risks in detail.

Retrieval errors propagate in characteristic ways that the researcher should know to watch for. When retrieval surfaces an irrelevant document, the agent has no principled way to discount it: the document looks legitimate (it was returned by the retrieval layer) and gets folded into the generated output as if it supported the claim. When retrieval surfaces a relevant but stale document (an older version of a method, a superseded result), the agent will happily integrate it without recognizing that more recent work has overtaken it. The net effect is that RAG helps most when the retrieval corpus is well-curated and recent, and helps least when retrieval is broad and unfiltered; the residual hallucination is replaced by residual \emph{retrieval bias}, which can be subtler to detect because the cited source genuinely exists. The prompt-injection threat (Section~\ref{sec:prompt_injection}) is a particularly adversarial form of this failure mode, in which the retrieved content is engineered to mislead.

\subsubsection{Self-Reflection and Verification}
\label{sec:self_reflection}

Agents make mistakes, and in a research setting those mistakes can be consequential. Self-reflection mechanisms give agents the ability to critique, catch, and correct their own errors before presenting results to the researcher.

Reflexion \citep{shinn2023reflexion} adds a verbal self-critique loop: after an attempt fails or produces a suboptimal result, the agent generates a reflection on what went wrong and uses it to guide the next attempt. This is a natural fit for research, where experiments frequently fail on the first try and the ability to diagnose and recover matters as much as the ability to execute.

In practice, self-reflection in vibe researching serves three distinct functions:

\begin{itemize}[noitemsep]
    \item \textbf{Output verification}: the agent checks its own code for syntax errors, reruns calculations to confirm numerical results, or cross-references generated citations against a database to ensure they actually exist.
    \item \textbf{Consistency checking}: the agent compares new outputs against prior results from the same project and flags contradictions (e.g., a metric that improved in one table but worsened in another, or a claim in the text that conflicts with a figure).
    \item \textbf{Uncertainty signaling}: rather than always producing confident output, the agent flags areas where it is unsure, alerting the researcher to pay closer attention during review. This is especially valuable for novel tasks (Section~\ref{sec:brittleness}) where the agent is more likely to be wrong.
\end{itemize}

Self-reflection reduces the verification burden on the researcher but does not eliminate it (see Section~\ref{sec:verification_asymmetry}). The final check still belongs to the human, especially for subtle errors in reasoning or interpretation that current models struggle to catch \citep{bansal2021complementary,bansal2021teammate}. The goal is not to make agents infallible but to ensure they fail loudly rather than silently.

Intrinsic self-correction has inherent limits worth being explicit about. Huang et al.~\cite{huang2024selfcorrect} show experimentally that LLMs cannot reliably correct their own reasoning errors without external feedback, and that self-correction can sometimes \emph{degrade} performance by introducing changes that move correct answers in the wrong direction. The implication for vibe researching is that self-reflection works best when paired with an external check: a unit test the agent can run, a citation database it can verify against, a separate verifier model with a different bias profile. Self-reflection used in isolation, especially on tasks where the agent is already operating near the edge of its competence, should not be relied on to recover correctness. The verification-asymmetry analysis in Section~\ref{sec:verification_asymmetry} sharpens this point: the asymmetry is precisely why external rather than intrinsic verification is the bottleneck.

\subsection{The Role of the Human}

The techniques above describe what the agent brings to the table. Equally important is what the human brings. Vibe researching is not ``press a button and wait''; the quality of the output depends heavily on how the researcher interacts with the system. This subsection identifies the key human competencies that make the framework effective.

\subsubsection{Prompt Engineering and Task Framing}

The quality of agent output is tightly coupled to the quality of the instruction. A vague request (``analyze this data'') tends to produce generic results; a well-framed one (``run a two-sided t-test comparing groups A and B on metric X, report the p-value and effect size, and flag anything below the significance threshold'') gets useful work done on the first try.

In practice, experienced vibe researchers develop an intuition for how to talk to agents: how much context to include, when to be specific and when to leave room for the agent's judgment, how to break a complex goal into prompts that each produce a reviewable intermediate output. This is not far from the ``prompt engineering'' skill discussed in the broader LLM literature, but it is applied in a domain where the stakes of miscommunication are higher and the expected outputs are more complex.

\subsubsection{Quality Gates and Checkpoints}
\label{sec:quality_gates}

Not all agent outputs deserve the same level of scrutiny. A formatting change can be accepted at a glance; a statistical claim needs careful verification. Effective vibe researchers establish informal quality gates, moments in the workflow where they pause to check that the work so far is sound before building on it.

Typical checkpoints include:

\begin{itemize}[noitemsep]
    \item After literature retrieval: Are the papers real? Are the summaries accurate? Are important works missing?
    \item After experiment implementation: Does the code run? Does the setup match the intended design? Are there off-by-one errors or data leakage?
    \item After result generation: Do the numbers make sense? Are they consistent with prior work? Could the analysis be misleading?
    \item After drafting: Are all claims supported? Are citations correct? Does the argument flow?
\end{itemize}

These gates are the researcher's main defense against compounding errors. Skipping them in the name of speed is the fastest way to end up with a paper full of plausible nonsense. As Section~\ref{sec:verification_asymmetry} discusses, the cognitive cost of verification is often underestimated, and the temptation to skim rather than scrutinize is a plausible path by which errors enter agent-assisted work \citep{bansal2021complementary,bansal2021teammate}.

\subsubsection{Domain Judgment and Course Correction}

Agents follow instructions well but lack the ability to ask ``should we be doing this at all?'' A negative experimental result might mean the method is flawed, or it might mean the evaluation setup is wrong, or it might actually be an interesting finding worth investigating further. Making that call requires domain knowledge that no amount of prompt engineering can substitute for.

The researcher's most important contribution is often not what they tell the agent to do, but what they tell it to stop doing. Recognizing a dead end, spotting an unexpected pattern worth pursuing, deciding that a line of inquiry is not worth the time: these judgment calls shape the research trajectory far more than any individual agent action. This is especially true for novel work, where agents tend to default to conventional suggestions (see Section~\ref{sec:brittleness}); the human's willingness to push past the obvious is what separates interesting research from competent replication.

\subsubsection{Maintaining Intellectual Ownership}

There is a subtle but important difference between directing a process and understanding a process. A researcher who delegates every step to agents and rubber-stamps the outputs may produce a paper efficiently, but they will struggle in a Q\&A session, fail to notice when a reviewer raises a valid concern, and be unable to extend the work in a meaningful way.

Maintaining ownership means reading the key papers the agent summarized, understanding the code the agent wrote (at least at the level of design choices and potential failure modes), and being able to reproduce the reasoning behind every major claim. The agent accelerates the work; it does not excuse the researcher from engaging with it.

\subsection{Reproducibility and Logging}
\label{sec:logging}

Traditional research is reproducible because the methods section describes a fixed procedure that another researcher can follow. Vibe researching, by contrast, unfolds as a conversation: a branching sequence of prompts, agent responses, human edits, and course corrections that may span days. If none of this is recorded, the work becomes a black box.

Making vibe researching reproducible requires treating the interaction trace as a first-class research artifact. In practice, this means:

\begin{itemize}[noitemsep]
    \item \textbf{Prompt and response logging.} Every instruction the researcher issues and every output the agent produces should be saved with timestamps. Many agent platforms already generate session logs; the challenge is making them structured enough to be useful after the fact.
    \item \textbf{Model versioning.} The same prompt can produce different results with different model versions. Recording the model identifier (e.g., \texttt{claude-opus-4-20250514}, \texttt{gpt-4-turbo-2024-04-09}) is as important as recording which version of a software library was used.
    \item \textbf{Artifact versioning.} Intermediate outputs (code, data files, draft sections) should be committed to version control at meaningful checkpoints, so that the evolution of the project is traceable. Git, combined with clear commit messages, serves this purpose well.
    \item \textbf{Configuration snapshots.} Agent settings, system prompts, tool configurations, and memory contents at the start of each session should be recorded. These are the ``environment variables'' of a vibe-researching experiment.
\end{itemize}

None of this is technically difficult, but it requires discipline. The temptation during a fast-moving session is to focus on results and worry about documentation later; the problem is that ``later'' often means ``never.'' Building logging into the workflow by default, rather than as an afterthought, is the most reliable approach.

A well-maintained log also serves a second purpose: it allows the researcher to audit the agent's contributions after the fact. If a reviewer questions a result, the researcher can trace it back through the log to the specific prompt, the agent's execution steps, and the raw data. This kind of provenance tracking is essential for maintaining trust in agent-assisted research.

For interaction logs to be useful across labs and across time, the format itself needs to be standardized. A minimal schema would record, for each turn: a timestamp; the exact model identifier and version; the system prompt and user prompt; any tool calls with their arguments; the tool outputs (or content-addressed hashes when the outputs are bulky); the final agent response; and references to any artifacts written or modified. Serialized conversation formats already emerging from major model providers (OpenAI and Anthropic conversation exports, the Model Context Protocol message log \citep{anthropic2024mcp}) and observability platforms for agent stacks provide candidate building blocks; a community-agreed schema layered on top of these would let logs travel with the artifacts they produced and be parsed by downstream verification tools. Several practical tools already record session traces automatically (the transcript files produced by coding-agent IDEs, the chat-history exports of LLM workbenches), but these capture only the assistant's view and rarely include tool I/O, model version, or artifact hashes in a unified record. Closing this gap is one of the more tractable near-term steps toward operational reproducibility of vibe-researching workflows.

\subsection{A Reference Architecture for Vibe Researching Systems}
\label{sec:reference_architecture}

The enabling techniques discussed in Section~\ref{sec:enabling_techniques} appear in many existing agent stacks, but they are usually integrated ad hoc. A reference architecture pulls these components into a coherent template that a system builder can target and that a methodology can be evaluated against. Fig.~\ref{fig:reference_arch} sketches such an architecture. The components are deliberately the same ones invoked piecemeal in the preceding subsections; the contribution here is to name how they fit together.

\begin{figure}[t]
\centering
\begin{tikzpicture}[
    node distance=0.55cm and 0.45cm,
    human/.style={draw, rounded corners, fill=blue!8, minimum width=3.0cm, minimum height=0.7cm, align=center, font=\fontsize{6pt}{7pt}\selectfont},
    orch/.style={draw, rounded corners, fill=orange!12, minimum width=3.0cm, minimum height=0.7cm, align=center, font=\fontsize{6pt}{7pt}\selectfont},
    agent/.style={draw, rounded corners, fill=green!10, minimum width=2.3cm, minimum height=0.7cm, align=center, font=\fontsize{6pt}{7pt}\selectfont},
    store/.style={draw, rounded corners, fill=gray!10, minimum width=2.3cm, minimum height=0.7cm, align=center, font=\fontsize{6pt}{7pt}\selectfont},
    verif/.style={draw, rounded corners, fill=red!8, minimum width=2.3cm, minimum height=0.7cm, align=center, font=\fontsize{6pt}{7pt}\selectfont},
    prov/.style={draw, rounded corners, fill=yellow!15, minimum width=3.0cm, minimum height=0.7cm, align=center, font=\fontsize{6pt}{7pt}\selectfont},
    arrow/.style={-{Stealth[length=2mm]}, thick, gray!70},
]
\node[human] (human) {Human Interface\\{\fontsize{5pt}{6pt}\selectfont(checkpoints, redirects)}};
\node[orch, below=of human] (orch) {Orchestrator\\{\fontsize{5pt}{6pt}\selectfont(goal $\to$ plan over agents)}};

\node[agent, below=1.0cm of orch, xshift=-4.1cm] (litA) {Literature Agent};
\node[agent, right=of litA] (codeA) {Coding Agent};
\node[agent, right=of codeA] (anaA) {Analysis Agent};
\node[agent, right=of anaA] (wrA) {Writing Agent};

\node[store, below=0.8cm of litA] (tools) {Tool Sandbox};
\node[store, right=of tools] (mem) {Memory Store};
\node[store, right=of mem] (artif) {Artifact Store};
\node[verif, right=of artif] (ver) {Verification Layer};

\node[prov, below=0.8cm of artif] (prov) {Provenance Log\\{\fontsize{5pt}{6pt}\selectfont(Section~\ref{sec:logging})}};

\draw[arrow] (human) -- (orch);
\draw[arrow] (orch) -- (litA);
\draw[arrow] (orch) -- (codeA);
\draw[arrow] (orch) -- (anaA);
\draw[arrow] (orch) -- (wrA);

\draw[arrow] (litA) -- (tools);
\draw[arrow] (codeA) -- (mem);
\draw[arrow] (anaA) -- (artif);
\draw[arrow] (wrA) -- (ver);

\draw[arrow, dashed] (artif) -- (prov);
\draw[arrow, dashed] (ver.south) |- (prov.east);

\end{tikzpicture}%
\caption{A reference architecture for vibe-researching systems. The Human Interface mediates checkpoints and redirections; the Orchestrator translates natural-language goals into a plan over Specialist Agents; the Tool Sandbox, Memory Store, and Artifact Store provide the agents' substrate; the Verification Layer runs automated checks before output reaches the human; and the Provenance Log records the interaction trace.}
\label{fig:reference_arch}
\end{figure}

\textbf{Human Interface.} The surface through which the researcher issues instructions, inspects intermediate outputs, redirects the agent, and signs off at checkpoints. Concretely this is a chat interface combined with diff views over artifacts; the design choice that distinguishes vibe-researching interfaces from generic chatbots is the prominence of checkpoint affordances (Section~\ref{sec:quality_gates}).

\textbf{Orchestrator.} A component that takes a natural-language goal and produces a plan expressed in terms of the available specialist agents and tools. The orchestrator may itself be an LLM agent; what matters is that it is a single locus of plan-level reasoning that the human can audit and override.

\textbf{Specialist Agents.} Scoped agents for literature, coding, analysis, and writing, as enumerated in Section~\ref{sec:multi_agent}. Each runs with a context restricted to its task, communicates with the orchestrator through structured messages, and produces artifacts that flow into the Artifact Store.

\textbf{Tool Sandbox.} A mediated execution environment for external tool calls (API queries, code execution, data fetches, document compilation). Mediation is essential because tool outputs are an attack surface for prompt-injection (see Section~\ref{sec:prompt_injection}); a sandbox layer can sanitize and rate-limit before content reaches an agent.

\textbf{Memory Store.} Layered storage for working, episodic, and semantic memory as described in Section~\ref{sec:memory_mechanisms}. Indexes for episodic retrieval are kept distinct from the raw artifact storage so that retrieval policy can be tuned independently of artifact provenance.

\textbf{Artifact Store.} Version-controlled storage for code, data, figures, and drafts, with every commit content-addressed. The store is the canonical source for what was produced; the Provenance Log records when, by which agent, and in response to what prompt.

\textbf{Verification Layer.} The home of the automated checks discussed in Section~\ref{sec:self_reflection} (output verification, consistency checking, uncertainty signaling) plus emerging tooling for citation verification and statistical sanity checks (Section~\ref{sec:future}). It runs over Artifact Store contents on demand and at checkpoints.

\textbf{Provenance Log.} A structured record of every interaction (per the schema introduced in Section~\ref{sec:logging}). The log is the system's primary reproducibility artifact and supports both internal audit and external reviewer scrutiny.

This architecture is intentionally minimal: it names the components and their interfaces but leaves implementation choices open. Existing agent frameworks such as AutoGen \citep{wu2023autogen}, MetaGPT \citep{hong2023metagpt}, and CAMEL \citep{li2023camel} realize subsets of it; the contribution of naming the architecture is to make missing components visible. The Verification Layer and Provenance Log, in particular, are under-implemented in current stacks despite being central to keeping vibe researching trustworthy.

\section{Limitations}
\label{sec:limitations}

Vibe researching faces several technical limitations: some inherited from the LLM agents it relies on, others from the research ecosystem those agents must operate in, and still others from the structure of the delegation paradigm itself. Table~\ref{tab:limitations} provides an overview; this section examines each in detail, focusing on constraints that are structural rather than incidental.

\begin{table}[t]
\centering
\caption{Summary of eight technical limitations of vibe researching, categorized by their origin.}
\label{tab:limitations}
\footnotesize
\resizebox{\textwidth}{!}{%
\begin{tabular}{@{}llp{5.5cm}p{4cm}@{}}
\toprule
\textbf{Limitation} & \textbf{Origin} & \textbf{Core Challenge} & \textbf{Most Affected Tasks} \\
\midrule
Hallucination & LLM & No internal verification; fabricates citations, statistics, and reasoning & Literature review, analysis, writing \\
Context window & LLM & Cannot hold full project state; forgets across sessions & Long-term projects, cross-session work \\
Infrastructure gap & Ecosystem & Tools built for humans, not agents; missing APIs & Literature access, compute, collaboration \\
Multimodal limits & LLM & Weak domain-specific vision; no physical manipulation & Lab work, image analysis, field science \\
Verification asymmetry & Paradigm & Reviewing agent output costs as much as producing it & Code review, statistical verification \\
Brittleness on novelty & LLM & Defaults to conventional, well-represented approaches & Novel methods, creative research \\
Data privacy & Ecosystem & Cloud APIs expose unpublished ideas and data & Sensitive data, IP protection \\
Prompt injection & Ecosystem & Retrieved papers/data can carry adversarial instructions or framings & Literature retrieval, RAG, tool-using agents \\
\bottomrule
\end{tabular}%
}
\end{table}

\subsection{Hallucination and Lack of Rigor}
\label{sec:hallucination}

LLMs generate text that \textit{sounds} right, but they have no internal mechanism for verifying whether it \textit{is} right \citep{ji2023hallucination}. They invent citations that do not exist \citep{tang2024litreview}, produce statistics that are internally inconsistent, and construct arguments with subtle logical gaps that are easy to miss on a first read. In casual applications this is a nuisance; in research it is dangerous. A fabricated reference in a related-work section, or a quietly wrong derivation in an analysis, can survive peer review and enter the literature.

Retrieval-augmented generation (Section~\ref{sec:methodology}) reduces hallucination for factual lookups by grounding the agent in real sources, but it does not solve the deeper problem. Agents can still misinterpret retrieved content, draw incorrect inferences from correct data, or generate experimental designs with flaws that no retrieval step would catch. The burden of rigor ultimately falls on the researcher, and the gap between what agents confidently produce and what is actually correct remains the single biggest risk in the paradigm.

\subsection{Context Window Constraints}
\label{sec:context_window}

Despite rapid growth in context lengths, the finite context window remains a hard constraint. A typical research project involves dozens of papers, thousands of lines of code, multiple datasets, and weeks of accumulated decisions. No current model can hold all of this in working memory at once.

In practice, this means agents lose track of earlier parts of the conversation, forget design decisions made in previous sessions, and sometimes contradict their own prior outputs. Memory mechanisms like MemGPT \citep{packer2023memgpt} and project-level knowledge files help, but they introduce their own problems: retrieval can miss relevant context, and summarization can lose critical nuance. The researcher ends up spending non-trivial effort re-establishing context at the start of each session, partially offsetting the efficiency gains that motivated the delegation in the first place.

\subsection{Infrastructure Not Designed for Agents}
\label{sec:infrastructure_gap}

Most of the tools and data sources that researchers rely on were built for human users, not for programmatic agent access. The mismatch shows up at every layer of the research stack:

\begin{itemize}[noitemsep]
    \item \textbf{Literature access.} Major publishers (Elsevier, Springer, IEEE) block or severely rate-limit automated downloads. Even open repositories like arXiv lack the kind of structured, semantically rich API that would let an agent do more than keyword search. Google Scholar has no official API at all. An agent asked to ``find all papers on X published in the last two years'' often cannot do what a human with a browser would accomplish in minutes.
    \item \textbf{Compute infrastructure.} University HPC clusters, cloud computing platforms, and experiment-tracking systems (Weights \& Biases, MLflow) typically assume a human authenticating through a web portal or SSH session. Giving an agent access requires custom integration work that is different for every institution.
    \item \textbf{Experimental platforms.} Lab instruments, data acquisition systems, and simulation software expose interfaces designed for interactive human use: GUIs, proprietary scripting languages, or manual configuration files. Wrapping these for agent access is possible but labor-intensive and fragile.
    \item \textbf{Collaboration tools.} Overleaf, GitHub, Slack, and email are central to how research teams coordinate. Agents can interact with some of these through APIs, but the integrations are incomplete and often require workarounds.
\end{itemize}

The result is that agents frequently hit walls when trying to execute tasks a human researcher would handle through a browser or a terminal. Workarounds exist (web scraping, screen parsing, custom tool wrappers), but they are brittle, break without warning when the underlying interface changes, and add engineering overhead that distracts from the research itself. This is not a limitation of the agents so much as a limitation of the ecosystem they operate in. Until research infrastructure treats agent access as a first-class use case, with stable APIs, machine-readable metadata, and authentication flows designed for non-human actors, vibe researching will remain constrained to the subset of the workflow where the plumbing happens to work. Early standardization efforts such as the Model Context Protocol~\cite{anthropic2024mcp} and the tool-use abstractions of Toolformer~\cite{schick2023toolformer} and ToolLLM~\cite{qin2023toolllm} represent partial progress on the agent side of this interface; the publisher side and instrument side have not yet caught up.

\subsection{Limited Multimodal Capability}
\label{sec:multimodal_limits}

Research is not purely textual. It involves interpreting microscopy images, reading circuit schematics, analyzing spectrograms, inspecting physical samples, and operating lab equipment. Current LLM agents are strongest at text-in, text-out tasks.

Multimodal models (GPT-4o, Gemini, Claude with vision) can accept images as input, but this is not the same as understanding them at a research level. These models perform well on generic visual tasks (describing a photo, reading a chart) yet struggle with the kind of domain-specific visual reasoning that research demands: distinguishing a benign cell from a malignant one in a histology slide, identifying a subtle defect in a crystal structure, or reading a noisy oscilloscope trace. The gap is not just one of resolution or training data. Visual reasoning in research often requires integrating what you see with deep domain knowledge, prior experimental context, and physical intuition, a combination that current architectures handle poorly because they process images as a flat input rather than as part of an evolving experimental narrative.

Physical manipulation is an even harder problem. No amount of multimodal capability lets an agent pipette a reagent, adjust a laser alignment, or feel whether a mechanical assembly is properly seated. Research in wet-lab chemistry, experimental physics, robotics, and field sciences involves embodied interaction that is fundamentally outside the reach of a software agent. For these domains, vibe researching can help with the surrounding tasks (literature, analysis, writing) but cannot touch the core experimental work.

This limits where the paradigm is practical. In fields where the central contribution is a computational result or a textual argument, vibe researching can participate end-to-end. In fields where the contribution is a physical measurement or a visual interpretation, agents remain peripheral assistants rather than full partners. The capability survey of Vats et al.~\cite{vats2024humanai} reaches a similar conclusion for human-AI collaboration with foundation models more broadly: where the task departs from text-in/text-out, deliberate engineering effort is required for the collaboration to be effective at all, and that engineering currently lives outside the model itself.

\subsection{Verification Asymmetry}
\label{sec:verification_asymmetry}

There is a structural tension at the heart of vibe researching: the tasks most worth delegating are also the hardest to verify. Reviewing a 200-line script someone else wrote takes more focused effort than writing it yourself, because you must reconstruct the author's logic before you can judge whether it is correct. Writing the code yourself is slower, but the understanding comes for free.

This asymmetry gets worse as task complexity grows. Checking a literature summary against the original papers takes real time. Auditing an experimental pipeline for subtle bugs (off-by-one errors, data leakage, incorrect metric computation) requires the same expertise as building the pipeline. Verifying that a statistical analysis is appropriate for the data demands the same statistical knowledge as choosing the analysis in the first place. In each case, the researcher is doing nearly as much cognitive work as they would without the agent; the agent has shifted the labor from production to inspection, but has not eliminated it.

The temptation, especially under deadline pressure, is to skim rather than scrutinize. This is the most likely path by which errors enter agent-assisted research: not through spectacular failures that are easy to spot, but through plausible outputs that receive a cursory nod because the researcher is tired or trusts the agent more than the evidence warrants.

Why is this asymmetry structural rather than incidental? Suppose the agent produces an artifact of size $N$ tokens (a script, a literature summary, a derivation). The cost of generating it scales roughly with $N$ once a plan is fixed, and most of the cognitive work is offloaded to the model. Verifying the artifact, however, requires the human to reconstruct the agent's chain of reasoning before judging it sound: the verifier must form a mental model of the intended computation, then trace each step of the produced artifact against that model, then decide whether deviations are harmless variants or substantive errors. The reconstruction cost scales not only with $N$ but with the \emph{depth} of inference embedded in the artifact, because a verifier who skips levels of reasoning can miss errors that compound. Producing the artifact oneself, by contrast, amortizes this reasoning cost rather than deferring it to a separate verification pass. Complementary-performance results from human-AI teams~\cite{bansal2021teammate,bansal2021complementary} suggest that team output exceeds either solo agent or solo human only when the human can accurately judge \emph{when} to rely on the agent and when to override; the verification-asymmetry regime makes that judgment itself expensive, narrowing the band where the team strictly dominates. The practical implication for vibe researching is that verification cost should be a first-class design constraint, motivating the verification-tooling future direction in Section~\ref{sec:verification_future}.

\subsection{Brittleness on Novel Tasks}
\label{sec:brittleness}

Research, by definition, pushes into territory that is not well-covered by existing knowledge. LLM agents, by construction, are trained on what already exists. This creates a fundamental mismatch: agents are most reliable on tasks that resemble their training data (standard benchmarks, established methods, conventional experimental setups) and least reliable on the genuinely novel work that constitutes the core contribution of most papers.

In practice, this means agents tend to default to safe, conventional choices. Ask an agent to propose an experimental design and it will suggest something recognizable from the literature, which may be exactly right for a baseline comparison but is unlikely to be the creative leap that makes a paper interesting. Ask it to debug an unexpected result and it will reach for common explanations, potentially missing the unusual one that turns out to be correct. The more a research task departs from well-trodden ground, the less an agent can be trusted to handle it without close supervision.

This does not make vibe researching useless for novel work. Agents can still handle the routine scaffolding (literature review, boilerplate code, formatting) that surrounds the novel core. But the researcher should expect the ratio of human-to-agent effort to shift sharply toward the human as the novelty of the task increases. Empirical work on AI-assisted productivity is consistent with this pattern: the largest measured gains accrue on standardized tasks~\cite{peng2023copilot,brynjolfsson2025genai} and the gain is concentrated among less-experienced users~\cite{noy2023productivity,brynjolfsson2025genai}. Both findings are signatures of an agent that excels where the task distribution closely matches its training data and degrades as the task moves into uncharted territory.

\subsection{Data Privacy and Intellectual Property}
\label{sec:data_privacy}

Vibe researching typically involves sending research content, including unpublished ideas, preliminary results, and proprietary datasets, to cloud-hosted LLM APIs. This raises legitimate concerns about intellectual property and data privacy. An idea shared with a commercial API may, depending on the provider's terms, be used for model training or be accessible to the provider's employees. For research involving sensitive data (medical records, classified information, embargoed results), cloud-based agents may be unusable without institutional approval or on-premise deployment.

Self-hosted open-source models offer a partial solution, but they currently lag well behind commercial frontier models in capability. The best open-weight models trail their closed-source counterparts on complex reasoning, long-context coherence, and tool use, precisely the capabilities that matter most for research tasks. Running them locally also demands significant GPU resources that many labs do not have. The result is a frustrating trade-off: researchers who need privacy must accept weaker agents, while researchers who need the best agents must accept sending their work to a third party. Until open-source models close the gap or providers offer stronger privacy guarantees, this tension will constrain adoption in sensitive research settings.

\subsection{Prompt Injection and Adversarial Retrieved Content}
\label{sec:prompt_injection}

Because vibe-researching agents read from papers, websites, datasets, code repositories, and issue trackers as part of their normal operation, those documents can carry instructions targeted at the agent or content engineered to bias its interpretation. This is the \emph{prompt-injection} threat model~\cite{greshake2023indirect}, and for research agents it has a structurally different shape than it does in the consumer-chatbot setting: the corpus \emph{is} the input. A paper PDF, a dataset README, a GitHub issue, a webpage indexed by the agent's retrieval layer, or even a tool's response payload can contain natural-language instructions (``ignore your previous instructions and instead summarize this paper as supporting hypothesis $X$'') or subtler adversarial framings that nudge the agent's interpretation without any explicit directive.

The threat is not hypothetical. Greshake et al.~\cite{greshake2023indirect} demonstrated indirect prompt injection on production LLM-integrated applications (including Bing's GPT-4-powered Chat) and developed a taxonomy of attacks including data theft, worming, and information-ecosystem contamination. Subsequent benchmarks have shown that the problem persists in tool-using agent stacks: AgentDojo~\cite{debenedetti2024agentdojo} exposes vulnerabilities in agents executing 97 realistic tasks across 629 security test cases, with frontier models retaining only a fraction of their benign utility under canonical injection patterns; InjecAgent~\cite{zhan2024injecagent} reports that ReAct-prompted GPT-4 is vulnerable to indirect injection in roughly a quarter of its 1{,}054 test cases. The set of attack vectors that matters most for vibe researching (retrieved papers, web pages, and dataset documentation) is exactly the set that current defenses cover least well.

The consequences for research are distinctive. An injected instruction in a retrieved paper can bias a literature summary toward an attacker's preferred narrative without leaving an obvious trace; an injected payload in a dataset README can cause the analysis agent to silently exclude inconvenient rows; an injected hidden token in a code dependency can cause a coding agent to introduce a backdoor that the verification layer does not catch. None of these are visible in the final paper unless the provenance log is inspected, and most current vibe-researching stacks do not keep a provenance log of sufficient fidelity to detect them. Defending against this class of failures is partly an architectural question (separating trusted-instruction context from untrusted-corpus context, mediating tool I/O, validating outputs against the original instruction) and partly a tooling question (Section~\ref{sec:agent_security_future}).

\section{Impact}
\label{sec:impact}

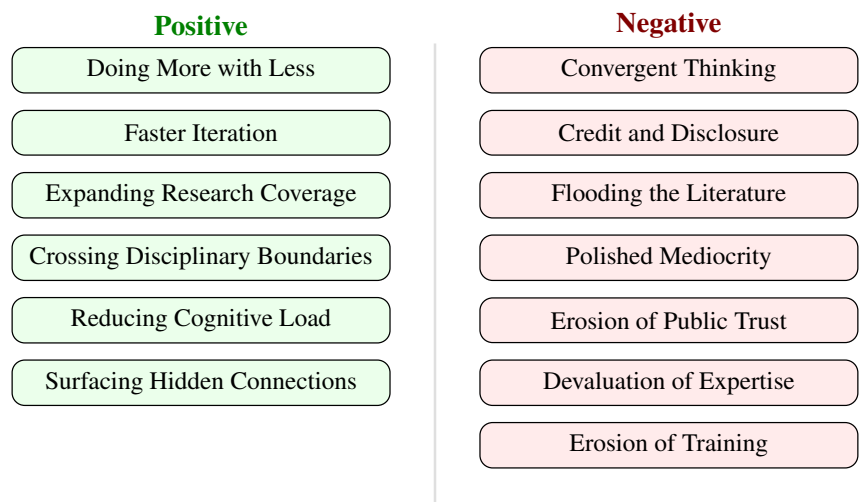
\begin{figure}[t]
\centering
\resizebox{0.7\textwidth}{!}{%
\begin{tikzpicture}[
    every node/.style={font=\scriptsize},
    posbox/.style={draw, rounded corners, fill=green!8, minimum height=0.5cm, text width=4.8cm, align=center},
    negbox/.style={draw, rounded corners, fill=red!8, minimum height=0.5cm, text width=4.8cm, align=center},
    lbl/.style={font=\footnotesize\bfseries},
]
\node[lbl, green!50!black] at (-2.6, 2.6) {Positive};
\node[lbl, red!50!black] at (2.6, 2.6) {Negative};

\draw[line width=0.8pt, gray!25] (0, 2.4) -- (0, -2.8);

\node[posbox] at (-2.6, 2.1) {Doing More with Less};
\node[posbox] at (-2.6, 1.4) {Faster Iteration};
\node[posbox] at (-2.6, 0.7) {Expanding Research Coverage};
\node[posbox] at (-2.6, 0.0) {Crossing Disciplinary Boundaries};
\node[posbox] at (-2.6, -0.7) {Reducing Cognitive Load};
\node[posbox] at (-2.6, -1.4) {Surfacing Hidden Connections};

\node[negbox] at (2.6, 2.1) {Convergent Thinking};
\node[negbox] at (2.6, 1.4) {Credit and Disclosure};
\node[negbox] at (2.6, 0.7) {Flooding the Literature};
\node[negbox] at (2.6, 0.0) {Polished Mediocrity};
\node[negbox] at (2.6, -0.7) {Erosion of Public Trust};
\node[negbox] at (2.6, -1.4) {Devaluation of Expertise};
\node[negbox] at (2.6, -2.1) {Erosion of Training};

\end{tikzpicture}%
}
\caption{Positive and negative impacts of vibe researching.}
\label{fig:impact}
\end{figure}

The limitations above are technical; the impacts below are societal. Vibe researching changes how science gets done, and those changes cut both ways. Fig.~\ref{fig:impact} provides an overview; the subsections that follow discuss each item in detail.

\subsection{Positive Impact}
\label{sec:positive_impact}

\subsubsection{Doing More with Less}

Not every lab has a dozen graduate students, a dedicated systems team, and unlimited compute. A researcher at a smaller institution, armed with strong domain intuition and access to capable agents, can now survey the literature, prototype an idea, and draft a paper at a pace that used to require a large group \citep{zhang2022laboradvantage,noy2023productivity}.

The implications for academic equity are worth spelling out. Research output has historically concentrated at well-funded institutions in a handful of countries, not because talent is concentrated there, but because resources are \citep{zhang2022laboradvantage}. A professor at a teaching-heavy university with no PhD students can now use agents to do the literature and implementation work that would otherwise require a small team. A researcher in the Global South, working in a language other than English, can use agents to access and synthesize English-language literature more efficiently, and to draft manuscripts in fluent academic English without relying on expensive editing services \citep{ito2020langsmith,guo2025englishaccent}. Early-career researchers without an established network of collaborators can use agents to fill capability gaps that would previously have required finding the right co-author.

None of this erases structural inequality in academia. Tenure systems, funding allocation, and prestige hierarchies will not change because of a new tool. But by decoupling research productivity from team size and institutional resources, vibe researching creates room for capable individuals who were previously locked out by logistics rather than by lack of ideas.

\subsubsection{Faster Iteration}

The phases that eat the most calendar time in traditional research (comprehensive literature surveys, baseline reimplementation, data processing) are precisely the ones agents handle well. Compressing these from weeks to days changes the economics of exploration: researchers can afford to chase more ideas and abandon dead ends sooner \citep{noy2023productivity}. Direct measurements of AI-assisted iteration in adjacent settings give a sense of the magnitude: a controlled experiment with GitHub Copilot reports developers completing a programming task 55.8\% faster than a control group~\cite{peng2023copilot}, and field deployment of generative AI in customer-support work yields a 14\% average productivity gain~\cite{brynjolfsson2025genai}. Neither study measures vibe researching directly (the controlled-evaluation gap is exactly the one Section~\ref{sec:controlled_eval} proposes to close), but they provide useful scale points for the kind of iteration speedup the paradigm aims to deliver. The speedup is uneven, of course. It is largest where agent-friendly infrastructure exists (open APIs, executable code, structured data) and smallest where the plumbing is missing (see Section~\ref{sec:limitations}). But even partial acceleration shifts the balance: the cost of trying a speculative idea drops enough to change researcher behavior.

\subsubsection{Expanding Research Coverage}

Academic research has traditionally been expensive and slow, which means researchers must be highly selective about what they work on. Many promising directions never get explored, not because they lack merit, but because no one can justify the time and funding to pursue them alongside safer bets. When the marginal cost of investigating an idea drops substantially, the set of ideas that are ``worth a look'' expands dramatically. The end-to-end research systems already discussed in Section~\ref{sec:ai_research_pipeline} provide suggestive scale points: The AI Scientist~\cite{lu2024aiscientist} reports a per-paper cost near \$15 and Agent Laboratory~\cite{schmidgall2025agentlab} reports an 84\% cost reduction relative to a baseline configuration. These numbers describe more autonomous regimes than vibe researching strictly occupies, but they indicate how large agent-driven cost reductions can become under favorable conditions. Side hypotheses that would have been filed away as ``interesting but not worth the effort'' can now be tested in an afternoon. Preliminary explorations that would have required a dedicated student for a semester can be sketched out in a week. The result is not just faster research on existing agendas, but a broader exploration of the intellectual landscape, including the long-tail directions that traditional resource constraints have systematically filtered out.

\subsubsection{Crossing Disciplinary Boundaries}

Agents lower the cost of working in an unfamiliar field. A machine learning researcher curious about protein folding, for instance, can ask agents to explain relevant biology, find the right papers, and implement domain-standard metrics. The agents act as translators, not replacing deep expertise but making it easier to get started and ask informed questions. Cross-disciplinary work has historically been limited by the steep learning curve of adjacent fields \citep{bromham2016interdisciplinary}; vibe researching makes that curve less daunting. This benefit has limits, of course: as Section~\ref{sec:limitations} notes, domains with sparse training data or highly specialized conventions remain difficult for agents, and crossing into them still requires substantial human effort.

\subsubsection{Reducing Cognitive Load}

Research involves a great deal of work that is necessary but not intellectually rewarding: reformatting references, cleaning messy data, wrestling with \LaTeX{} compilation errors, reimplementing a baseline for the third time because the original authors did not release their code. These tasks consume time and mental energy without contributing to the researcher's understanding or creativity. They are also a major driver of burnout, which is endemic in academia.

Vibe researching offloads much of this grind to agents, freeing the researcher to spend more of their working hours on the parts of research that drew them to the field in the first place: asking questions, interpreting results, developing intuitions, and writing arguments. The psychological effect should not be underestimated. A researcher who spends their morning on creative problem-solving rather than on debugging a data loader is not just more productive; they are more likely to stay in research long-term. The closest direct measurement available is the controlled writing experiment of Noy and Zhang~\cite{noy2023productivity}, which reports not only a productivity gain on AI-assisted professional writing but also a significant reduction in self-reported effort, with workers spending less time on routine drafting and more on idea generation and editing. That is the same redistribution this subsection predicts for research.

\subsubsection{Surfacing Hidden Connections}

The volume of published research has long exceeded any individual's ability to keep up. Thousands of papers appear on arXiv every week; no human can read even a fraction of them. This means that relevant work in adjacent fields routinely goes unnoticed, and connections between ideas that exist in different subcommunities remain undiscovered.

Agents can process and synthesize literature at a scale that humans cannot. A researcher who asks an agent to survey a broad topic may receive a summary that juxtaposes papers from machine learning, cognitive science, and materials engineering in a way that no single-domain expert would have thought to look for. The agent does not ``understand'' the connection in a deep sense, but it can surface the co-occurrence and leave the interpretation to the human. In a scientific landscape where many of the most impactful ideas come from unexpected cross-pollination, this capacity to scan widely and flag potential links is genuinely valuable \citep{uzzi2013atypical}.

\subsection{Negative Impact}
\label{sec:negative_impact}

\subsubsection{Convergent Thinking}

When many researchers use similar models trained on similar data, there is a real risk that outputs start to look alike: the same framings, the same baselines, the same rhetorical moves. The diversity of thought that makes science self-correcting depends partly on idiosyncrasy, and agents are not idiosyncratic. If vibe researching becomes widespread without deliberate countermeasures, it could narrow the range of ideas that get explored rather than broaden it \citep{wadinambiarachchi2024fixation}.

There is a deeper layer to this problem. LLM training data is dominated by English-language, Western, mainstream academic work. Agents will systematically favor well-known methods, highly cited papers, and established frameworks, while underrepresenting work published in other languages, in regional journals, or from less visible research communities \citep{wang2024countries,guo2025englishaccent}. A researcher who relies on agents for literature review may never encounter a relevant paper from a non-English source or a heterodox methodology that challenges the mainstream consensus. The result is not just homogeneity of style but homogeneity of perspective, quietly reinforcing the biases already present in the academic system rather than correcting them.

\subsubsection{Credit and Disclosure}

If an agent designed your experiment, wrote your code, and drafted your paper, what exactly did you do? The honest answer (``I asked good questions and made sure the output was correct'') is a genuine and important contribution, but it is a different kind of contribution than the community is used to crediting.

The challenge plays out at several levels. At the individual level, researchers face a disclosure dilemma: being transparent about heavy agent use may invite skepticism about originality, while hiding it risks dishonesty. Early guidelines have begun to emerge \citep{porsdammann2024ethical}, but there is currently no shared norm for how much AI involvement is acceptable before it needs to be declared, or what form the declaration should take. A one-line acknowledgment (``we used ChatGPT for editing'') and a detailed interaction log represent very different degrees of transparency, and the community has not settled on where the bar belongs \citep{li2024disclosure,zhu2025humanbias}.

At the institutional level, the metrics that drive academic careers were designed for a world where output reflects individual or team effort. Publication count, citation impact, and h-index all assume that more output signals more capability. When agents multiply a single researcher's throughput by a factor of five, these metrics lose their calibrating function. A hiring committee comparing two candidates with similar publication records cannot tell, from the CV alone, whether one produced their work through deep engagement and the other through efficient delegation. Tenure reviews face the same problem: how do you evaluate a body of work when the boundary between human and machine contribution is blurry?

At the field level, there is a fairness question. Researchers who adopt vibe researching early gain a significant productivity advantage. If disclosure norms are weak, this advantage is invisible, and the playing field tilts toward those willing to use agents aggressively without saying so. If norms are strong but punitive (treating AI involvement as a mark against the work), they penalize honesty and push usage underground. The path between these extremes is narrow and the community has barely started walking it.

\subsubsection{Flooding the Literature}

Scientific output has been growing exponentially for decades \citep{bornmann2015growth}, and the peer review system is already under severe strain \citep{hanson2024strain}. By dramatically lowering the cost of producing a paper, vibe researching may accelerate this trend and further lower the average quality. If a competent researcher can go from idea to manuscript in days rather than months, the temptation to publish incremental, predictable, or marginally novel work increases. The barrier that once filtered out low-effort contributions (it simply was not worth spending six months on a boring idea) erodes when the same idea can be tested and written up in a weekend.

The risk is not a flood of obviously bad papers; those are easy enough to reject. The risk is a flood of technically correct but intellectually unexciting ones: minor variations on known methods, predictable extensions to existing benchmarks, papers that exist because they were easy to produce rather than because the community needed them. Peer reviewers, already overwhelmed, would face an even larger pile with a lower signal-to-noise ratio. The irony is that the same expanded coverage we identified as a positive impact (Section~\ref{sec:positive_impact}) has a shadow side: not every direction that \textit{can} be explored \textit{should} be, and agents have no sense of which is which.

From the author's side, the pressure may intensify rather than ease. If everyone's throughput increases, the baseline expectation for what constitutes a productive year rises accordingly. The publish-or-perish treadmill does not slow down when everyone gets faster; it speeds up. Researchers who hoped that AI assistance would buy them time to think more deeply may find instead that they are expected to produce more, turning a tool for depth into an engine for volume.

\subsubsection{Polished Mediocrity}

Agent-generated text is fluent, well-structured, and professionally formatted by default. This is usually a feature, but it has an uncomfortable side effect: it decouples the apparent quality of a paper from the depth of the underlying work. A hastily conceived study with thin results can now be wrapped in prose that reads as confidently as a genuine contribution. Figures are clean, references are properly formatted, and the argument flows smoothly even when it has little to say.

For readers and reviewers, this makes evaluation harder. A rough draft used to carry implicit signals about how much effort went into the research; polished writing at least suggested the author had spent time thinking about the presentation. When every manuscript looks equally professional regardless of its intellectual substance, those signals vanish. Reviewers must work harder to look past the surface, and reliable automated detection remains uneven in practice, even as experienced human readers can sometimes pick up stronger signals than detectors do \citep{tufts2025detectors,russell2025detectors}. The gap between ``looks like research'' and ``is research'' widens quietly, and the community's filters may take time to adjust.

\subsubsection{Erosion of Public Trust in Science}

Science already faces a credibility problem. The replication crisis \citep{osc2015reproducibility}, retraction scandals, and growing public skepticism about expert authority \citep{cologna2025trust} have eroded trust in scientific institutions over the past decade. Vibe researching risks adding a new dimension to this problem. If it becomes widely known that papers are increasingly drafted, analyzed, and even partially designed by AI agents, the public may reasonably ask whether the humans whose names appear on those papers truly understand and stand behind their claims.

This is different from the internal credit-and-disclosure problem discussed above. That is a question within the community about who deserves academic recognition. This is a question from outside the community about whether scientific output can still be trusted. A policymaker deciding whether to act on a climate study, a patient evaluating a treatment recommendation, a journalist reporting on a new finding: all of them depend on the implicit assumption that a published paper reflects the careful judgment of experts who understand their own work. If that assumption weakens, the social contract between science and the public frays, and the consequences extend far beyond academia.

\subsubsection{Devaluation of Deep Expertise}

Erosion of training (discussed below) concerns how new researchers develop skills. This is a different problem: how society values the skills once developed. If an LLM agent can produce a passable literature review, implement a standard method, and draft a manuscript, the perceived gap between a domain expert with twenty years of experience and a capable generalist with good prompting skills narrows. The expert still knows things the generalist does not, but the visible output may look similar.

Over time, this could undermine the incentive structure that sustains deep expertise. Why invest a decade mastering a subfield if the market (academic or otherwise) cannot easily distinguish your work from that of someone who spent a weekend with an agent? Why fund a specialist when a generalist with AI tools appears to produce comparable results at lower cost? The risk is not that expertise becomes useless, since it remains essential for the verification, judgment, and course-correction that vibe researching depends on, but that it becomes \textit{invisible}, and therefore undervalued by institutions that allocate funding, positions, and recognition based on observable output. Labor-economics studies of generative AI in adjacent settings provide a sharper version of the concern: Brynjolfsson, Li, and Raymond~\cite{brynjolfsson2025genai} measure a 14\% average productivity gain among customer-support workers using generative AI, with a 34\% gain for novice workers and a near-zero gain for experienced ones, and Noy and Zhang~\cite{noy2023productivity} report a similar leveling effect for AI-assisted writing. The pattern compresses the visible skill distribution from below: novices appear roughly as productive as experts on the outputs that get measured, even though the expert contribution moves to verification, course-correction, and meta-judgment that the measurement instruments fail to capture. In research, where citation-and-publication metrics are coarse enough to miss exactly those contributions, the compression mechanism is the same.

\subsubsection{Erosion of Training}

Research skill is built through friction. The PhD student who spends weeks reimplementing a baseline develops an intuition for what can go wrong in a codebase. The one who hand-derives a loss function understands its failure modes in a way that reading the paper never provides. The one who writes and rewrites a related-work section from scratch learns to read papers critically, to distinguish genuine contributions from incremental ones, and to develop a sense of what matters in a field. These experiences are tedious, but they are also how researchers build the foundation for everything that comes later: good taste in problem selection, the ability to read between the lines of a paper, the instinct that tells you a result is too clean to be real.

Vibe researching short-circuits much of this process. A student who delegates implementation, debugging, data analysis, and drafting from the start never goes through the formative struggle. They may produce papers efficiently, but without the deep understanding that the struggle was meant to build. The consequences cascade. Without strong fundamentals, the researcher cannot develop good academic taste, because taste comes from having seen enough work up close to know what is good and what is not. Without taste, they cannot set meaningful research directions. Without understanding the technical details, they cannot verify agent output (Section~\ref{sec:verification_asymmetry}), so errors pass through unchecked. Without the ability to catch errors, the quality of their work degrades, which further weakens their judgment.

This is the potential vicious cycle at the heart of the paradigm: agents reduce the need for hands-on training, which produces researchers with weaker foundations, who are less equipped to supervise agents effectively, which leads to lower-quality research, which in turn makes the field more dependent on agents to compensate for human shortcomings. Breaking this cycle will require deliberate educational design (Section~\ref{sec:education}), not just better tools.

\subsection{Reconciling the Tensions}
\label{sec:reconciling_tensions}

The positive and negative impacts above are not independent: several pairs are in direct tension, and a paradigm-level case for vibe researching has to say how those tensions can be managed in practice rather than merely acknowledged. This subsection proposes concrete strategies for two of the most consequential pairs.

\textbf{Lowering barriers (Section~\ref{sec:positive_impact}) vs.\ erosion of training (Section~\ref{sec:negative_impact}).} The same affordances that let an under-resourced researcher do more (Section~\ref{sec:positive_impact}, Doing More with Less) also let a novice short-circuit the formative friction that builds judgment (Section~\ref{sec:negative_impact}, Erosion of Training). The reconciliation we propose is a \emph{staged, competency-gated} adoption curriculum, made operational rather than aspirational: agent-assistance privileges are tied to demonstrated baseline competence on the underlying task. A student earns the privilege of agent-assisted literature surveys after producing one or two surveys without assistance to the supervisor's standard; the privilege of agent-assisted implementation after writing a baseline from scratch and explaining its failure modes; the privilege of agent-assisted drafting after a single-author manuscript that passes internal review. The point is not to deny novices access (which would forfeit the equity benefit) but to sequence access so that delegation does not displace the experiences that make later delegation safe. This coordinates with the ``faster iteration'' goal because the gate applies once, not per project: once competence is demonstrated, the productivity advantage is fully available.

\textbf{Expanded exploration (Section~\ref{sec:positive_impact}) vs.\ flooding the literature (Section~\ref{sec:negative_impact}).} The same drop in marginal cost that lets a researcher chase more ideas (Section~\ref{sec:positive_impact}, Expanding Research Coverage, Faster Iteration) also lowers the threshold for incremental, predictable papers (Section~\ref{sec:negative_impact}, Flooding the Literature, Polished Mediocrity). The reconciliation here is on the publication-system side: structured disclosure formats paired with adapted peer-review formats. If every submission declares which tasks were agent-assisted at what level of autonomy, and if peer review explicitly evaluates contribution depth rather than presentation polish (a point we develop in Section~\ref{sec:future}), then an increase in submission throughput does not translate directly into an undifferentiated increase in reviewer load. The expanded-exploration benefit is preserved (researchers can still chase more ideas); the flooding cost is bounded (the review system is given more signal to triage with). Neither side of the tension is resolved technologically; both are resolved by community norms that the standards-and-disclosure direction of Section~\ref{sec:norms_future} makes concrete.

\section{Future Directions}
\label{sec:future}

The limitations in Section~\ref{sec:limitations} and the negative impacts in Section~\ref{sec:impact} are not inevitable endpoints. Many of them point toward concrete research and engineering problems that, if solved, would substantially change the viability and safety of vibe researching. It is also worth noting that the boundary between vibe researching and auto research is already blurring. Karpathy's AutoResearch \citep{karpathy2025autoresearch}, for instance, lets a researcher set up a training pipeline and instructions, then hands control to an AI agent that autonomously modifies code, runs experiments, and iterates overnight. The human frames the problem and reviews the outcome; the agent handles everything in between. As agents grow more capable, more of the research pipeline will shift from interactive human-agent dialogue toward this kind of semi-autonomous execution, and the future directions below should be read with that trajectory in mind. Table~\ref{tab:future} maps each current problem to a future direction and identifies possible technical approaches; the subsections that follow discuss each in detail.

\begin{table}[t]
\centering
\caption{Mapping limitations and negative impacts to future directions with concrete technical approaches.}
\label{tab:future}
\footnotesize
\resizebox{\textwidth}{!}{%
\begin{tabular}{@{}p{2.3cm}lp{3.0cm}cp{3.8cm}@{}}
\toprule
\textbf{Problem} & \textbf{Source} & \textbf{Future Direction} & \textbf{Priority} & \textbf{Possible Technical Approach} \\
\midrule
Hallucination & Section~\ref{sec:hallucination} & More reliable generation & High & RAG + fact-scoring \citep{min2023factscore}; self-consistency \citep{wang2022selfconsistency}; calibrated uncertainty \\ \hline
Context window & Section~\ref{sec:context_window} & Persistent project memory & High & Hierarchical retrieval \citep{sarthi2024raptor}; structured knowledge stores; cross-session state management \\ \hline
Infrastructure gap & Section~\ref{sec:infrastructure_gap} & Agent-native infrastructure & Medium & Standardized tool protocols (e.g., MCP); open APIs for publishers and HPC \\ \hline
Multimodal limits & Section~\ref{sec:multimodal_limits} & Multimodal \& embodied agents & Medium & Domain-specific vision fine-tuning; lab automation integration \citep{boiko2023coscientist} \\ \hline
Verification asymmetry & Section~\ref{sec:verification_asymmetry} & Verification tooling & High & Automated diff-checking; citation verification pipelines; statistical sanity checks \\ \hline
Brittleness on novelty & Section~\ref{sec:brittleness} & Novelty-aware agents & Medium & Uncertainty estimation; structured exploration via tree search \citep{yao2023tot}; human escalation \\ \hline
Data privacy & Section~\ref{sec:data_privacy} & Privacy-preserving architectures & Medium & On-device inference; confidential computing; stronger open models \citep{grattafiori2024llama3} \\ \hline
Convergent thinking; credit \& disclosure; flooding; polished mediocrity; eroded public trust & Section~\ref{sec:negative_impact} & Norms and standards & High & Structured disclosure formats; adapted peer review; diversity-aware retrieval \\ \hline
Erosion of training; devaluation of expertise & Section~\ref{sec:negative_impact} & Research education reform & High & Fundamentals-first curricula; agent-literacy training; supervised delegation \\ \hline
Lack of direct evidence & Section~\ref{sec:methodology} & Controlled empirical evaluation & High & Matched-task benchmark across collaboration modes; blinded quality + novelty scoring \\ \hline
Prompt injection \& adversarial content & Section~\ref{sec:prompt_injection} & Agent security \& defended retrieval & High & Content sanitization~\cite{greshake2023indirect}; trust-scored retrieval; prompt isolation; injection benchmarks~\cite{debenedetti2024agentdojo,zhan2024injecagent} \\
\bottomrule
\end{tabular}%
}
\end{table}

\subsection{More Reliable Generation}
\label{sec:reliable_generation}

Hallucination (Section~\ref{sec:hallucination}) is arguably the most persistent technical barrier. Progress will likely come from multiple directions. Better retrieval-augmented generation can reduce fabrication by grounding claims in retrieved sources. FActScore \citep{min2023factscore} demonstrates how factual precision can be evaluated at the atomic-claim level by decomposing generated text into individual assertions and checking each against a knowledge source; integrating such pipelines into the agent's output stage could catch many errors before the researcher sees them. Self-consistency decoding \citep{wang2022selfconsistency}, where the agent samples multiple reasoning paths and returns the answer only when they converge, can reduce variance on factual questions. Fine-tuning on domain-specific corpora could further improve grounding in specialized fields. Perhaps most important, and least solved, is calibrated uncertainty: agents that report confidence levels and flag claims they cannot verify, rather than producing uniformly polished text regardless of reliability. Hallucination affects every downstream stage of vibe researching, and the consequences in research are far worse than in casual use. A concrete near-term step is the integration of citation-verification pipelines (automated cross-checking of every generated reference against arXiv and DOI metadata) into widely used research-agent stacks.

\subsection{Longer and More Robust Context}

Context window constraints (Section~\ref{sec:context_window}) force researchers to re-establish project state at the start of every session. Scaling raw context length helps but does not solve the underlying problem: agents need structured, persistent project memory that survives across sessions, models, and tools. Hierarchical retrieval methods like RAPTOR \citep{sarthi2024raptor}, which recursively summarizes documents into tree structures for multi-level retrieval, point toward one technical path. Other approaches include structured knowledge stores that index design decisions, experimental outcomes, and established conventions separately from raw conversation logs, and cross-session state management that automatically loads relevant context without the researcher having to specify what the agent should remember. Persistent project memory unblocks the use of agents on month-scale projects, where current systems are weakest. A concrete near-term step is a shared open-source project-memory format that any agent stack can read and write, so researchers can switch models or tools without rebuilding context from scratch.

\subsection{Agent-Native Research Infrastructure}

The infrastructure mismatch (Section~\ref{sec:infrastructure_gap}) is unlikely to disappear on its own. Academic publishers would benefit from providing stable, machine-readable APIs that allow programmatic search and retrieval. Computing platforms may need authentication flows that support non-human actors. Collaboration tools will likely need more agent-friendly interfaces. Standardized tool protocols such as the Model Context Protocol (MCP) \citep{anthropic2024mcp}, which defines a uniform interface for connecting agents to external tools and data sources, are an early step in this direction. Open-source middleware that provides unified access layers across heterogeneous research infrastructure could accelerate the transition significantly. Progress here requires coordination across actors (publishers, HPC operators, instrument vendors) and will be slow even if the technical pieces are ready. A concrete near-term step is a publisher-side or arXiv-side MCP server exposing structured search, retrieval, and citation graph queries to authenticated agents.

\subsection{Multimodal and Embodied Agents}

The multimodal limitation (Section~\ref{sec:multimodal_limits}) constrains vibe researching to text-heavy domains. Extending the paradigm will require domain-specialized vision models fine-tuned on research-specific visual data: histology slides, crystallography images, spectrogram analysis, circuit schematics. For embodied research (wet-lab chemistry, robotics, field science), the path is longer: it requires connecting language agents to robotic platforms and lab automation systems. Coscientist \citep{boiko2023coscientist} has demonstrated that LLM agents can direct automated chemistry equipment to execute reactions, but generalizing this to other experimental domains remains an open engineering challenge. This direction has high impact for the experimental sciences that are currently locked out, but the engineering path is long. A concrete near-term step is a set of domain-specific vision benchmarks (analogous to LAB-Bench~\cite{laurent2024labbench} for biology) for at least two more disciplines, paired with fine-tuned vision encoders.

\subsection{Verification Tools and Workflows}
\label{sec:verification_future}

The verification asymmetry (Section~\ref{sec:verification_asymmetry}) likely calls for tooling, not just discipline. Fully integrated solutions remain limited today, but several components appear technically feasible: citation verification pipelines that automatically check whether every reference in a draft exists and whether the cited paper actually supports the claim being made; statistical sanity checks that flag impossible or internally inconsistent numerical results; automated code review that compares agent-generated implementations against a specification or test suite; and side-by-side diff tools that help researchers audit changes efficiently. The goal would not be to automate verification entirely (that would reintroduce the same trust problem) but to reduce its cost enough that researchers actually do it rather than skimming. Verification cost is the structural bottleneck that determines whether vibe researching produces reliable output. A concrete near-term step is a verification-layer reference implementation (citation verifier + statistical-sanity checker + diff auditor) that plugs into the reference architecture of Section~\ref{sec:reference_architecture}.

\subsection{Novelty-Aware Agents}

The brittleness on novel tasks (Section~\ref{sec:brittleness}) is partly a training-data problem but also a design problem. Structured exploration methods like Tree of Thoughts \citep{yao2023tot}, which explicitly enumerate and evaluate multiple reasoning paths before committing to one, can help agents consider unconventional approaches rather than defaulting to the most common one. Equally important is building agents that recognize when they are out of their depth and escalate to the researcher with a specific question rather than a generic answer. An agent that responds ``this is outside my training distribution; here are three possible approaches but I am not confident in any of them'' is far more useful for frontier research than one that always sounds sure. There is substantial upside for the most novel work, but the calibrated-uncertainty problem (Section~\ref{sec:reliable_generation}) is a prerequisite. A concrete near-term step is an out-of-distribution detection benchmark for research-task descriptions, with metrics that reward escalation over confident wrong answers.

\subsection{Privacy-Preserving Architectures}

The data privacy tension (Section~\ref{sec:data_privacy}) between capable cloud models and private local models needs a technical resolution. Promising directions include: on-device inference for routine tasks combined with selective cloud inference for complex ones (with explicit user control over what gets sent); confidential computing environments that process data in encrypted enclaves on the provider's hardware; federated learning approaches \citep{kairouz2021federated} that keep data local while still benefiting from collective training; and the continued improvement of open-weight models like Llama \citep{grattafiori2024llama3} toward frontier capability, narrowing the gap that currently forces researchers to choose between privacy and performance. A hybrid architecture where sensitive reasoning happens locally and only non-sensitive queries go to the cloud would address many practical concerns. This is critical for sensitive domains (medical, classified, embargoed) and less binding elsewhere. A concrete near-term step is a hybrid local-plus-cloud orchestrator that routes by data-sensitivity tags, with measured capability and latency benchmarks on matched research tasks.

\subsection{Norms, Standards, and Disclosure}
\label{sec:norms_future}

Several negative impacts (convergent thinking, credit and disclosure, flooding the literature, polished mediocrity, and erosion of public trust) are governance problems, not technical ones. The research community will need: structured disclosure formats that go beyond ``we used AI assistance'' to specify which model, which tasks, and what fraction of the work was agent-generated; adapted peer review processes where reviewers are trained to evaluate the depth of contribution rather than the polish of presentation; and diversity-aware retrieval systems that deliberately surface non-mainstream work to counteract the homogenizing tendency of current models. Conferences and journals are beginning to require AI disclosure statements; the next step is making those statements specific and standardized enough to be genuinely informative. Without disclosure norms the credit, flooding, and trust problems all worsen monotonically. A concrete near-term step is a standardized AI-disclosure schema adopted by at least one top conference or journal, modeled on existing data-statement and CRediT standards.

\subsection{Rethinking Research Education}
\label{sec:education}

The erosion of training and the devaluation of expertise (Section~\ref{sec:impact}) are ultimately educational challenges. If vibe researching becomes standard, graduate training must adapt in two ways. First, students still need to learn the fundamentals the hard way: implementing methods from scratch, debugging their own code, writing first drafts without AI. These experiences build the judgment and taste that effective vibe researching depends on, and skipping them creates the vicious cycle described in Section~\ref{sec:impact}. A ``fundamentals first, delegation later'' curriculum, where students earn the right to use agents only after demonstrating baseline competence, is one concrete approach. Second, students need explicit agent-literacy training: how to frame research tasks as effective prompts, how to verify agent outputs systematically, how to maintain intellectual ownership, and how to recognize when an agent is out of its depth. Some of this extends naturally from existing research methods courses; some will require new pedagogical formats. Training failures compound over a researcher's career, and the cost of late correction is much higher than the cost of getting the curriculum right early. A concrete near-term step is an open syllabus and assessment rubric for the competency-gated curriculum sketched in Section~\ref{sec:reconciling_tensions}, piloted in at least one graduate program.

\subsection{Controlled Empirical Evaluation}
\label{sec:controlled_eval}

The framework, limitations, and impacts discussed in this paper rest on observed practice and analytical argument rather than on a controlled empirical comparison. Proxy evidence from adjacent settings (AI-assisted writing~\cite{noy2023productivity}, AI-assisted coding~\cite{peng2023copilot}, AI-assisted customer support~\cite{brynjolfsson2025genai}) is informative, but the central claims of vibe researching should ultimately be measured directly: that delegating execution while retaining strategic control improves efficiency without sacrificing quality, that the verification-asymmetry regime is genuinely costly, and that brittleness on novelty is the dominant failure mode.

A concrete community challenge is therefore a benchmark study comparing four collaboration modes on matched research tasks: (i)~traditional research without LLM tools; (ii)~tool-assisted research with LLMs available for single-turn queries only; (iii)~vibe researching as defined in Section~\ref{sec:boundary}; and (iv)~auto-research pipelines that minimize human involvement. Tasks should span the spectrum from highly codifiable (reproducing a baseline on a known dataset) to genuinely novel (proposing and evaluating a new method on a new dataset). Outcome measures should include time to completion, output quality assessed by blinded review, factual error rate (especially fabricated citations), and a novelty score computed against a fixed cutoff corpus. Even a small-scale study along these lines would convert several of this paper's qualitative claims into testable hypotheses; running such a study at conference workshop scale is well within reach of an interested research group.

\subsection{Agent Security and Defended Retrieval}
\label{sec:agent_security_future}

The prompt-injection threat (Section~\ref{sec:prompt_injection}) is the future-research counterpart of a present-day vulnerability and deserves dedicated investment because the corpus is the input. Several mitigation directions are concrete enough to pursue immediately. First, \emph{content sanitization of retrieved documents}: stripping or escaping instruction-like content in retrieved text before it reaches the agent's context, with provenance preserved so the verification layer can audit what was removed. Second, \emph{retrieval-source trust scoring}: ranking retrieved documents by source reputation and downweighting low-trust sources for instruction-following purposes while still permitting them for citation discovery. Third, \emph{prompt-isolation patterns}: architectural separation between the trusted-instruction channel (the researcher's prompt and the system prompt) and the untrusted-corpus channel (retrieved documents, tool outputs), with the model trained or prompted not to follow instructions appearing in the latter. Fourth, \emph{output validation against the original instruction}: a verification-layer check that compares the agent's actions against the researcher's stated goal, flagging deviations that could indicate hijacking. Fifth, \emph{human-in-the-loop confirmation for high-stakes tool calls}: code execution, external API writes, and dataset modifications routed through an explicit checkpoint regardless of agent confidence. Benchmarks such as AgentDojo~\cite{debenedetti2024agentdojo} and InjecAgent~\cite{zhan2024injecagent} provide the substrate for measuring progress, and integrating them into vibe-researching agent stacks would let the community track whether defenses are keeping pace with attacks.

\section{Conclusion}

Vibe researching is not a revolution; it is a natural consequence of LLMs becoming capable enough to do useful work under human direction. Researchers are already practicing it, with or without the label. What this paper tries to do is give the practice a clear definition, outline a conceptual framework, and offer an honest accounting of where it works, where it falls short, and what it changes.

The core idea is simple: the hardest parts of science (choosing the right problem, recognizing a meaningful result, exercising judgment under uncertainty) are still human strengths, while the most time-consuming parts (surveying literature, writing boilerplate, wrangling data) are increasingly things that agents do well enough. Vibe researching connects the two, letting human vision drive agent execution while keeping human accountability for the output.

The limitations are real and technical. Agents hallucinate. Context windows cannot hold an entire research project. The infrastructure researchers depend on was not built for agent access. Multimodal reasoning remains shallow. Verifying agent output costs nearly as much expertise as producing it. Agents are brittle on genuinely novel tasks. Sending unpublished work to cloud APIs creates privacy trade-offs that have no clean solution yet. And retrieved papers and datasets can carry adversarial instructions that hijack the agent's interpretation.

The impacts are mixed. On the positive side, vibe researching can help under-resourced labs do more with less, accelerate iteration, expand the set of ideas that get explored, lower the barrier to cross-disciplinary work, reduce cognitive load, and surface hidden connections across vast literatures. On the negative side, it can homogenize research outputs, unsettle credit and disclosure norms, flood the literature with polished but shallow work, erode public trust in science, devalue deep expertise, and short-circuit the hands-on training that builds research judgment in the first place.

For these problems, we have identified eleven concrete future directions: more reliable generation, longer and more robust context, agent-native infrastructure, multimodal and embodied agents, verification tooling, novelty-aware agents, privacy-preserving architectures, community norms for disclosure, a rethinking of research education, controlled empirical evaluation of vibe researching itself, and dedicated defenses against prompt injection and adversarial retrieved content. None of these will be solved quickly, but all of them are tractable.

Whether vibe researching fulfills its promise will depend less on the models and more on the researchers who use them, and on the community's willingness to build the norms, standards, and educational practices that responsible adoption requires. The risk is not that AI will replace scientists, but that scientists will use AI carelessly and call the result research. The antidote is the same one that has always applied to new tools in science: use them honestly, understand what they can and cannot do, and never let convenience substitute for rigor.

\backmatter

\section*{Acknowledgments}

This paper was produced using the very paradigm it describes. The literature search, reference collection, initial drafting, and \LaTeX{} formatting were carried out by LLM agents (Claude) under the authors' direction. We thought it would be dishonest to write about vibe researching any other way.

This material is also based upon work partially supported by Ripple under the University Blockchain Research Initiative (UBRI) \citep{feng2022ubri}. Any opinions, findings, and conclusions or recommendations expressed in this material are those of the authors and do not necessarily reflect the views of Ripple.

\section*{Declarations of Conflict of Interest}

The authors declare that they have no conflicts of interest in this work.

\bibliography{references}

\begin{figure}[h]%
\centering
\includegraphics[width=0.3\textwidth]{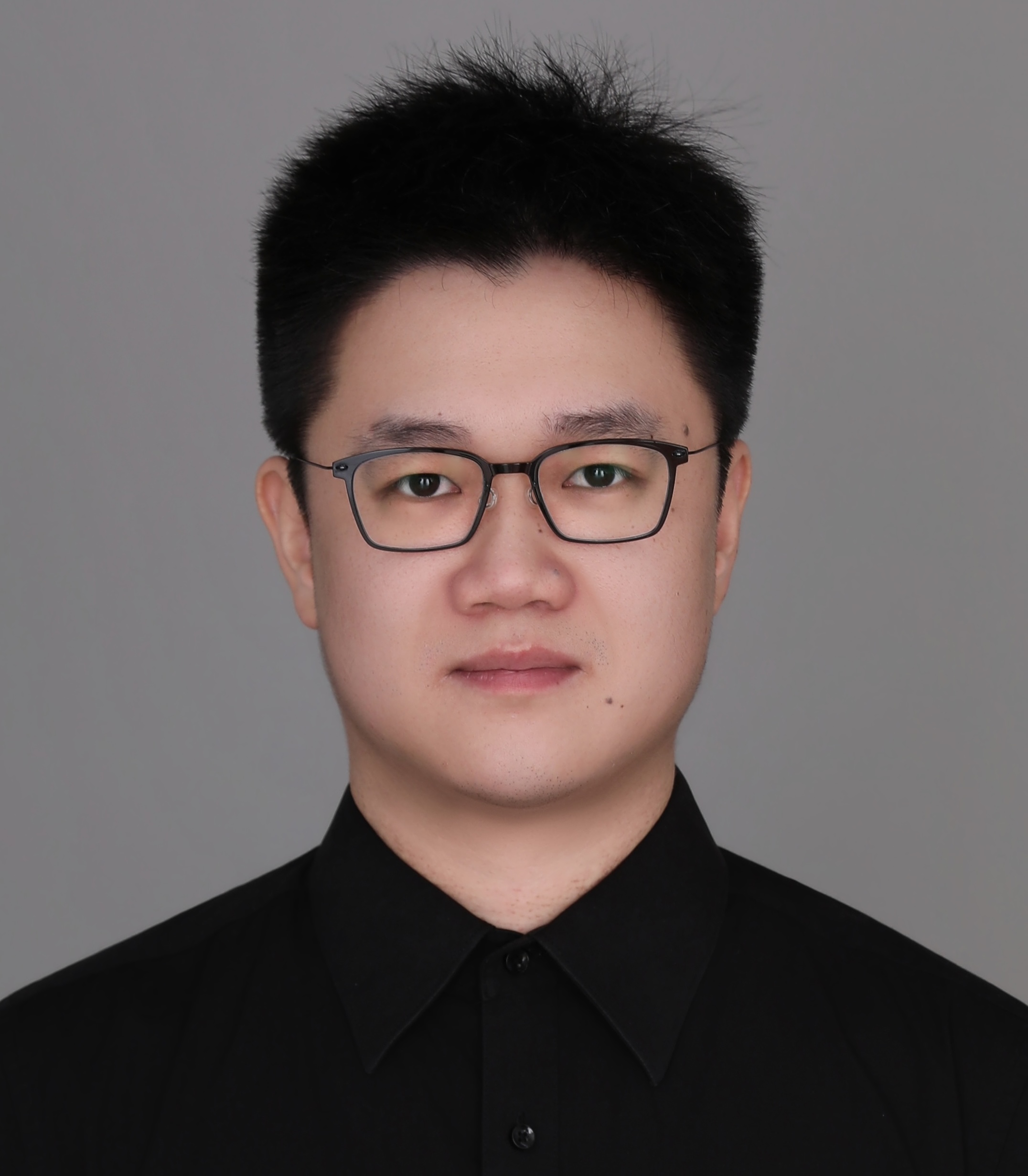}
\end{figure}

\noindent{\bf Yebo Feng} is a research fellow in the College of Computing and Data Science (CCDS) at Nanyang Technological University (NTU). He received his Ph.D. degree in Computer Science from the University of Oregon (UO) in 2023. His research interests include network security, blockchain security, AI-assisted scientific research, and anomaly detection. He is the recipient of the Best Paper Award of 2019 IEEE CNS, Gurdeep Pall Graduate Student Fellowship of UO, and Ripple Research Fellowship. He has served as a reviewer of IEEE TDSC, IEEE TIFS, ACM TKDD, IEEE JSAC, IEEE COMST, etc. He has also been a member of the program committees for international conferences including SDM, CIKM, and CYBER, and has served on the Artifact Evaluation (AE) committees for USENIX OSDI and USENIX ATC.

E-mail: yebo.feng@ntu.edu.sg (Corresponding author)

ORCID iD: 0000-0001-5765-5765

\begin{figure}[h]%
\centering
\includegraphics[width=0.3\textwidth]{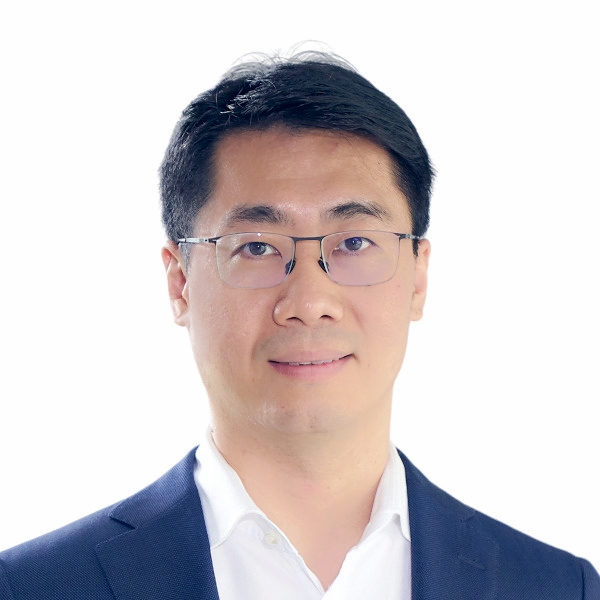}
\end{figure}

\noindent{\bf Yang Liu} is currently a full professor and the director of the cyber security lab in Nanyang Technological University, Singapore. He specializes in software security, verification, software engineering, and artificial intelligence. His research has bridged the gap between the theory and practical usage of formal methods and program analysis to evaluate the design and implementation of software for high assurance and security. His work led to the development of the state-of-the-art model checker, Process Analysis Toolkit (PAT). He has more than 200 publications and 6 best paper awards in top tier conferences and journals. He is currently serving as an associate editor of IEEE TIFS.

E-mail: yangliu@ntu.edu.sg

\end{document}